\documentclass[reprint,superscriptaddress,nofootinbib,amsmath,amssymb,aps,prd]{revtex4-1}
\usepackage[T1]{fontenc}
\usepackage{aecompl}
\usepackage{graphicx}
\usepackage{amsfonts}
\usepackage{amsmath}
\usepackage{bm}
\usepackage{tensor}
\usepackage{ulem}
\usepackage{mathrsfs}
\usepackage{rotating} 
\usepackage[varg]{txfonts}
\usepackage{fixltx2e} 
\usepackage{url}
\usepackage{etoolbox}
\usepackage{color}
\definecolor{MyGreen}{rgb}{0.0,.5,0.0}
\usepackage{hyperref}
 \hypersetup{
    colorlinks,%
    citecolor=MyGreen,%
    filecolor=blue,%
    linkcolor=blue,%
    urlcolor=blue
}
\providecommand{\adsurl}[1]{}

\providecommand\apj{ApJ}                 
\providecommand\aap{A\&A}            
\providecommand\mnras{MNRAS}

\providecommand\prd{Phys. Rev. D}

\providecommand\prl{PRL}

\providecommand\jcap{JCAP}

\providecommand\grg{Gen.~Rel.~Grav.}

\providecommand\annrevnucpartphys{Ann.~Rev.~Nucl.~Part.~Sci.}

\providecommand\cqg{Class.~Quant.~Grav.}   %

\newcommand{\CD}{\mathcal D}

\newcommand{\CQ}{\mathcal Q}

\usepackage{comment}

\newcommand{\rd}{{\mathrm d}}
\newcommand{\ddt}[1]{\frac{{\rm d}#1}{{\rm d}t}}
\newcommand{\caverage}[1]{\left\langle {\strut #1}\right\rangle_{\bar{\CD}}}
\newcommand{\inI}{{\rm I}}
\newcommand{\inII}{{\rm II}}

 3

\def\bea{\begin{eqnarray}}
\def\eea{\end{eqnarray}}

\definecolor{MyBloodyRed}{rgb}{0.7,0,0}

\definecolor{MyGreen}{rgb}{0.0,.5,0.0}

\definecolor{MyBlue}{rgb}{0.0,.0,1.0}

\begin{document}

\title[Dynamical volume partitioning of Newtonian cosmologies]{Cosmological Backreaction and the Separate Universe Conjecture:
\\Dynamical volume partitioning of Newtonian cosmologies}

\author{Thomas Buchert}
\email{buchert@cosmunix.de}
\affiliation{Universit\'e Lyon 1, ENS de Lyon, CNRS, CRAL, UMR 5574, Lyon, France}%
\author{Jan J. Ostrowski}
\email{Jan.Jakub.Ostrowski@ncbj.gov.pl}
\affiliation{Department of Fundamental Research, National Centre for Nuclear Research, Pasteura 7, 02-093 Warsaw, Poland}
\author{Alfred Bovon}
\email{alfred.bovon@ens-lyon.fr}
\affiliation{Universit\'e Lyon 1, ENS de Lyon, CNRS, CRAL, UMR 5574, Lyon, France}

\begin{abstract}
We provide a description of general Newtonian cosmologies for collisionless matter in terms of an exact and dynamical volume partitioning of global realizations of Newtonian models. Employing the formalism for the description of cosmological backreaction, an interaction between the inhomogeneities of the formation of structures in the Universe and their average properties can be analyzed. Our results can be regarded as a tool for many potential applications in the context of Newtonian simulations, but they also provide an exact way to assume analytical properties of the universe models on regional scales in conformity with global integral constraints on their average properties. We discuss approaches in relation to `separate Universe conjectures' in the literature and examine their consistency with the exact partitioning framework. In particular, using the Average Expansion Rate Approximation (AvERA) as an example, we show that identifying locally assigned Friedmannian rates with physical regional expansion rates is constrained by the global Newtonian $3$-torus condition. Alternatively, when the simulation cells are interpreted as non-material control volumes, Reynolds transport introduces mass flux and flux-backreaction terms, so that their dynamics is not in general that of independent Friedmann domains. We also discuss the corresponding consistency conditions for stable clustering prescriptions.

\end{abstract}
\maketitle

\textbf{Keywords:}
Newtonian cosmology -- volume partitioning -- separate Universe conjecture

\section{Introduction} \label{s-intro}

The cosmological backreaction connects the formation of structures and the expansion history of the Universe \cite{Buchert08status}. In order to quantitatively examine this connection, we investigate an exact Lagrangian volume partitioning of the universe model into spatial regions at different scales, e.g. for the two-scale problem into overdense ('virialized') and underdense (void) regions. In general relativity the former are characterized by
positive spatial scalar curvature and the latter by negative spatial scalar curvature. Due to the volume dominance of the latter regions, starting at the onset of structure formation, together with the dynamical non-conservation of spatial curvature \cite{buchertcarfora}, the emergence of negative scalar curvature on average is expected leading to a dark energy-like effect \cite{BuchRas12}.

In this work we concentrate on inhomogeneous Newtonian cosmologies with the motivation to establish a controlled analytical way of making assumptions on some regional scale of the problem, and with the perspective of providing tools for a corresponding general-relativistic treatment. The main reason to start within Newtonian cosmology is that we can globally control cosmological backreaction: vanishing spatial curvature together with the global topological architecture of a $3-$torus ensures vanishing of the kinematical backreaction term
that is necessary and sufficient for vanishing averages of the deviations from a global background cosmology
\cite{BuchertEhlers97}. Backreaction might be globally non-vanishing only for non-Euclidean background spacetimes and non-Euclidean global topologies.

In Newtonian cosmology, essentially two ingredients are needed to render the equations of Newtonian gravity unique: (i) introducing a global background and (ii) subjecting deviations thereof to periodic boundary conditions. This agrees with the common architecture of a Newtonian simulation in cosmology. 
Exact average properties can then be subjected to a global constraint, investigated in the fluid
approximation for a 'dust matter model' in the mean-field
approximation of self-gravitating collisionless matter in \cite{BuchertEhlers97} (see also \cite{EhlersBuchert97}). The Vlasov-Poisson dynamics of the mean-field approximation is briefly presented in appendix~\ref{appA}, and the Buchert-Ehlers theorem is reviewed in appendix~\ref{appB}.
We will then consider a multi-scale partitioning of the global Newtonian model, based on previous work in the framework of general relativity \cite{buchertcarfora} and \cite{multiscale}.

The result (\ref{general}) reviewed in appendix~\ref{appB} 
is valid for the average of the matter model `dust' on all compact domains from an infinitesimal domain up to any scale, e.g., up to the boundary of a compact and boundary-free universe model. 
We will in what follows include velocity dispersion that has been shown to not violate this theorem in \cite{Buchert18} (see also the discussion by \cite{Kaiser17}). We so extend the results to a general collisionless matter model.
Deviations from mean-field gravity are not considered in this paper but would furnish an interesting generalization possibility along the lines of the explicit coarse-graining investigated in \cite{Buchert05adhesive},
or along the lines of the BBGKY hierarchy advanced in \cite{gilles}, or by including noise \cite{mercader}, \cite{Lapi}.
Our results are presented for scalar variables; an extension to vector- and tensor-valued variables is straightforward within Newtonian cosmology, based on the seminal paper on the tensor virial conjecture by \cite{ChandraLee}, since averaging of tensors is well-defined in Euclidean space.

In section~\ref{s-foundations}, in order to provide a self-contained presentation, we briefly lay down the formalism of self-gravitating collisionless systems to establish the foundations on which this paper is based. 
Readers familiar with these foundations may skip this section including appendix \ref{appA} and move to section~\ref{s-partitioning}, where we partition the system to obtain an exact description of underdense and overdense regions and their nonlinear dynamical interaction, and we consider physical conditions to close this family of interacting systems of partitioning equations. We then discuss relations to `separate Universe conjectures' in the literature in section~\ref{sec:separate}, and we end with a discussion in section~\ref{s-discussion}. Appendices are devoted to velocity moments of Vlasov's equation and to known theorems and their generalization to include velocity dispersion.

\section{The Euler-Newton-Jeans system and spatial averages of its trace parts}
\label{s-foundations}
We consider a Galilei-Newton spacetime, foliated into flat spatial hypersurfaces $t=const.$, labeled by the Newtonian absolute time coordinate $t$. We also consider a non-rotational, inertial coordinate system in this spacetime with Euclidean geometry, realized by Cartesian coordinates $x_i  \ i =1,2,3$.

\subsection{Self-gravitating collisionless systems in Newtonian gravity}

In this section, we consider a continuous self-gravitating and collisionless system within Newtonian gravity.
The governing equations are re-derived in appendix \ref{appA}. We start by presenting the Euler-Newton-Jeans system, discuss its trace parts and the equation of state we obtain. 

\subsubsection{The Euler-Newton-Jeans system}

In Appendix \ref{appA}, we develop the calculations to obtain the first members of a hierarchy of velocity moment equations of Vlasov's equation that governs the dynamics of a self-gravitating collisionless continuum in Newtonian gravity in the mean-field approximation. These first members of the hierarchy are composed of the continuity equation for the mass density field $\varrho (x_i ,t)$, the Euler-Jeans equation for the mean velocity field components $\bar{v}_i (x_i , t)$, the evolution equation for the reduced second moment tensor components $\Pi_{ij} (x_i ,t)$, and the gravitational field equations for the gravitational field strength components in the mean-field approximation ${\bar g}_i (x_i , t)$:\footnote{A comma abbreviates the derivative with respect to the Eulerian coordinates $x_k$, but we often write out the derivative operator in what follows. We use the convention of summation over double indices. We keep the component and coordinate indices down, since we work in an Euclidean metric. }
\begin{equation}\label{ENJ1}
    \begin{aligned}
 \frac{\rd}{\rd t}\varrho &= - \varrho \bar{v}_{j,j}~;\\
 \frac{\rd}{\rd t}\bar{v}_i & = {\bar g}_i  - \frac{1}{\varrho} \Pi_{ij,j}~;\\
\frac{\mathrm{d}}{\mathrm{d}t} \Pi_{ij} &= -L_{ijk,k} - [\bar{v}_{i,k}\Pi_{jk} + \bar{v}_{j,k}\Pi_{ki} + \bar{v}_{k,k}\Pi_{ij}]~;\\ 
{\bar g}_{j,i} - {\bar g}_{i,j} &= 0 ~;\\
{\bar g}_{k,k} &= \Lambda - 4 \pi G \varrho~.
    \end{aligned}
    \end{equation}
In this system, $\rd / \rd t \equiv \partial / \partial t + \boldsymbol{\bar v} \cdot \boldsymbol{\nabla}$ denotes the Lagrangian derivative operator. The field equations can be rewritten in terms of the gravitational potential $\Phi$ defined by ${\bar g}_i =: - \Phi, i$, obeying the Poisson equation:
\begin{equation}
\Delta \Phi = 4\pi G \varrho - \Lambda \ .
\end{equation}
This system of equations is not closed unless the divergence of the third reduced velocity moment tensor $L_{ijk}$ is neglected, or its relation to other variables is given. We will call the above system the \textit{Euler-Newton-Jeans system} when dropping the third reduced velocity moment. 

\subsubsection{Discussion of the trace parts and thoughts on the equation of state}

The Euler-Jeans equation can be rewritten as:
\begin{equation}
\frac{\partial}{\partial t}\bar{v}_i + \bar{v}_k\frac{\partial}{\partial x_k}\bar{v}_i = {\bar g}_i - \frac{1}{\varrho}\frac{\partial}{\partial x_j}\Pi_{ij}
~. \end{equation}
We introduce a short-hand notation for the divergence of the symmetric tensor $ \boldsymbol{\Pi}$:
\begin{equation}
\Psi_i \equiv \frac{\partial}{\partial x_j}\Pi_{ij}
~. \end{equation}
We perform a spatial derivative with respect to $x_j$ and we take the trace of the equation to obtain:
\begin{equation}
\label{RAY}
\frac{\mathrm{d}}{\mathrm{d}t}\frac{\partial \bar{v}_i}{\partial x_i} + \frac{\partial\bar{v}_k}{\partial x_i}\frac{\partial \bar{v}_i}{\partial x_k}= \frac{\partial {\bar g}_i}{\partial x_i}-\frac{\partial }{\partial x_i}\Big( \frac{1}{\varrho}\Psi_i \Big)
~. \end{equation}
We then rewrite this equation in terms of kinematical invariants, splitting the velocity gradient into a symmetric and an antisymmetric part:
\begin{equation}
\frac{\partial {\bar v}_i}{\partial x_j} = \frac{1}{2} \Big(\frac{\partial {\bar v}_i}{\partial x_j} + \frac{\partial {\bar v}_j}{\partial x_i}\Big) + \frac{1}{2} \Big(\frac{\partial {\bar v}_i}{\partial x_j} - \frac{\partial {\bar v}_j}{\partial x_i}\Big) = \frac{1}{3}\theta\delta_{ij} + \sigma_{ij} + \omega_{ij}
~, \end{equation}
where in the second equality we introduced $\theta$ representing the trace (the rate of expansion), $\sigma_{ij}$ a symmetric traceless tensor (the shear), and $\omega_{ij}$ an antisymmetric tensor (the vorticity).
We furthermore define the rate of shear $\sigma$ and the rate of vorticity $\omega$ as follows:
\begin{equation}
\sigma^2 \equiv \frac{1}{2}\sigma_{ij}\sigma_{ij} \quad ; \quad \omega^2 \equiv \frac{1}{2}\omega_{ij}\omega_{ij}
~.
\end{equation}
We then define the two of the scalar principal invariants of the mean velocity gradient:
\begin{subequations}\label{invariants}
\begin{align}
\inI &= \boldsymbol{\nabla} \cdot \boldsymbol{\bar v} = \frac{\partial {\bar v}_i}{\partial x_i} = \theta ~; \\
\inII &= \frac{1}{2}\boldsymbol{\nabla}\cdot (\boldsymbol{\bar v}(\boldsymbol{\nabla} \cdot \boldsymbol{\bar v})-(\boldsymbol{\bar v}\cdot\boldsymbol{\nabla})\boldsymbol{\bar v} ) \nonumber\\
&= \frac{1}{2}\Big(\frac{\partial {\bar v}_i}{\partial x_i}\frac{\partial {\bar v}_j}{\partial x_j} - \frac{\partial {\bar v}_i}{\partial x_j}\frac{\partial {\bar v}_j}{\partial x_i}\Big) = \frac{1}{3}\theta^2 + \omega^2 - \sigma^2
~. \end{align}
\end{subequations}
Equation \eqref{RAY} then takes the form of \textit{Raychaudhuri's equation}:
\begin{equation}\label{Raychaudloc}
\frac{\mathrm{d}}{\mathrm{d}t}\theta = \Lambda -4\pi G \varrho-\frac{1}{3}\theta^2 + 2(\omega^2 - \sigma^2) - \frac{\partial }{\partial x_i}\Big( \frac{1}{\varrho}\Psi_i \Big)
~. \end{equation}
If we assume that typical velocities of the system stay in the vicinity of the mean velocity, $|\boldsymbol{v} - \bar{\boldsymbol{v}}| \simeq \varepsilon \bar{\boldsymbol{v}}$, we can estimate the moments of the hierarchy:
\begin{subequations}\begin{align}\label{anisPI1}
    \Pi_{ij} &\sim \varrho \varepsilon^2|\boldsymbol{\bar{v}}|^2 ~; \\
    L_{ijk} &\sim \varrho\varepsilon^3|\boldsymbol{\bar{v}}|^3 ~,
\end{align}
\end{subequations}
and may neglect the reduced third moment $L_{ijk}$ compared to terms $ \propto \varepsilon^2$ to restrict the problem to the \textit{Euler-Newton-Jeans system} (\ref{ENJ1}) with $L_{ijk,k} =0$. This argument is locally not realistic as a closure of the hierarchy, since in situations like shell-crossing or virialization all the moments are of comparable order. 
However, each member of the hierarchy is exact, and upon spatial averaging over a compact $3-$torus universe model, divergences become boundary terms that vanish for the empty-boundary $3-$torus model. This implies a global closure when averaging the reduced second moment equation (or, similarly, higher moment equations by going deeper into the hierarchy).

Locally, we have to resort to phenomenological closure assumptions, e.g. by `masking' the velocity ellipsoid by an isotropic filter function and consider an effectively 
isotropic `dynamical pressure' with an anisotropic distribution:
\begin{equation}\label{anisPI}
    \Pi_{ij} = p\delta_{ij} + \Delta_{ij} ~~;~~\textrm{with}~~ |\Delta_{ij}| \ll p ~.
\end{equation}
Under the approximation of vanishing (or averaged out) anisotropic contribution, the Euler-Newton-Jeans system (\ref{ENJ1}) with $L_{ijk,k} =0$ (which, strictly, is dynamically consistent only for $\sigma_{ij} =0$), reduces to the following: 
\begin{equation}\label{ENJ3}
    \begin{aligned}
 \frac{\rd}{\rd t}\varrho &= - \varrho \bar{v}_{j,j}~;\\
 \frac{\rd}{\rd t}\bar{v}_i & = {\bar g}_i  - \frac{1}{\varrho} p_{,i}~;\\
\frac{\mathrm{d}}{\mathrm{d}t} p\delta_{ij} &=- p[\bar{v}_{i,j} + \bar{v}_{j,i} + \bar{v}_{k,k}\delta_{ij}]~;\\ 
{\bar g}_{j,i} - {\bar g}_{i,j} &= 0 ~;\\
{\bar g}_{k,k} &= \Lambda - 4 \pi G \varrho~.
    \end{aligned}
    \end{equation}
This system has the form of hydrodynamical equations with pressure
for a self-gravitating continuum. However, it does not follow thermodynamical laws. Instead, the evolution equation for the isotropic reduced second moment tensor provides:
\begin{subequations}
\begin{equation}\label{ContinuityENJ}
    \Big( \frac{1}{p}\frac{\rd p}{\rd t} - \frac{1}{\varrho}\frac{\rd \varrho}{\rd t}\Big)\, \delta_{ij} = - (\bar{v}_{i,j} + \bar{v}_{j,i}) ~;
\end{equation}
in terms of the shear tensor we obtain:
\begin{equation}\label{ContinuityENJsig}
     \Big( \frac{1}{p}\frac{\rd p}{\rd t} - \frac{5}{3\varrho}\frac{\rd \varrho}{\rd t}\Big)\, \delta_{ij} = - 2\sigma_{ij} ~.
\end{equation}
Taking the trace and the trace-free part on both sides, we obtain the two independent equations:
\begin{equation}{\label{continuityENJsig0}}
    \frac{1}{p}\frac{\rd p}{\rd t} = \frac{5}{3}\frac{1}{\varrho}\frac{\rd \varrho}{\rd t}~~\textrm{and}~~\sigma_{ij} = 0 ~,
\end{equation}
\end{subequations}
i.e. the shear tensor is pointwise vanishing, which is only exact in the case of homogeneous-isotropic systems, discussed in the next section. Let us finally integrate, using Lagrangian coordinates $\bf X$, the previous equation for the dynamical pressure to yield:
\begin{equation}\label{eqstate1}
p (\boldsymbol{X},t) = \kappa(\boldsymbol{X})\varrho^{5/3}(\boldsymbol{X},t) ~~\textrm{with}~~ \kappa(\boldsymbol{X}) \equiv \frac{p(\boldsymbol{X},t_0)}{\varrho^{5/3}(\boldsymbol{X},t_0)} ~.
\end{equation}
The above integral is not an equation of state because $\kappa$ depends on the position $\boldsymbol{X}$. By fixing the initial conditions in such a way that the relation between p and $\varrho$ is initially the same (at each fluid element), we obtain an equation of state that is characteristic for the whole fluid and that is valid for each position and fluid element in Eulerian space. 
On phenomenological grounds, we may consider the more general \textit{polytropic equation of state},
\begin{equation}\label{eqstate2}
p = \kappa \varrho^{\gamma} ~~;~~ \kappa = \textrm{const.} ~~;~~ \gamma = \frac{n+1}{n} ~,
\end{equation}
where $n = const.$ is named the \textit{polytropic index}. Our analysis gives us $ \gamma = 5/3$ and $n=3/2$, but we can imagine that additional physics would alter this polytropic index (for more details, the reader may consult the review by \cite{Buchert05adhesive}).

\subsection{Self-gravitating collisionless systems in Newtonian cosmology}

In this section, we introduce a background cosmology plus deviations thereof in order to describe the dynamics with respect to a chosen background cosmology which we assume to be a member of the class of 
FLRW (Friedmann-Lema\^\i tre-Robertson-Walker) cosmologies. We closely follow  \cite{Peebles1980}. (For notational ease, we henceforth drop the overbar notation for the mean velocity field and the mean-field gravitational field strength.)

\subsubsection{Homogeneous-isotropic systems}

We start with a homogeneous deformation. A trajectory field ${\bf x} = {\bf f}({\bf X},t)$ then assumes the form: 
\begin{equation}\label{traj1}
    (\boldsymbol{f}_{H} )_i = a_{ij}(t)X_j~,
\end{equation}
with ${\bf X}$ denoting Lagrangian coordinates. Hence, the background universe model is homogeneous with $\varrho = \varrho_H(t)$. The continuity equation as the first member of the Euler-Newton-Jeans system, implies that the velocity is a linear function of $\boldsymbol{x}$, $(\boldsymbol{v}_H)_i = H_{ij}(t)x_j$. But, we also have as the primary assumption that the universe model is locally isotropic, meaning that the deformation matrix $\boldsymbol{a}$ reduces to $a_{ij}(t) = a(t)\delta_{ij}$ and that the fields can be written:
\begin{subequations}
\begin{equation}\label{fieldiso}
\boldsymbol{v}_H = H\boldsymbol{x} ~~;~~\boldsymbol{g}_H = \frac{\ddot{a}}{a}\boldsymbol{x} ~~;~~\textrm{and}~~\varrho_H(t) =\varrho_{H_0} a(t)^{-3} ~,
\end{equation}
with the \textit{Hubble function} $H(t) \equiv \dot{a}(t) / a(t)$.
Inserting this into the Euler-Newton-Jeans system we obtain:
\begin{equation}\label{Fried1}
    3\frac{\ddot{a}}{a} + 4\pi G\varrho_H - \Lambda = 0  ~,
\end{equation}
which is the Newtonian acceleration law for the scale factor $a(t)$.
To obtain the Newtonian expansion law, we integrate \eqref{Fried1} formally, by multiplying by $2a\dot{a}$ to obtain (with $k$ an integration constant):
\begin{equation}\label{Fried2}
    3\left(\frac{\dot{a}}{a}\right)^2 + \frac{3k}{a^2}- 8\pi G\varrho_H - \Lambda = 0 ~.
\end{equation}
\end{subequations}
We can write the Euler-Newton-Jeans system for homogeneous-isotropic cosmologies as follows:
\begin{equation}\label{ENJFRied1}
    \begin{aligned}
3 \frac{\ddot{a}}{a} + 4 \pi G\varrho_H - \Lambda &= 0 ~;\\
3 H^2 + \frac{3k}{a^2} - 8\pi G\varrho_H - \Lambda  &= 0 ~;\\ 
 \dot{\varrho}_H +3H\varrho_H &= 0 ~~;~~\varrho_H = \varrho_{H_0} a^{-3}~;\\
\frac{1}{p_H}\dot{p}_H = \frac{5}{3} \frac{1}{\varrho_H} \dot{\varrho}_H ~~;~~p_H&=\kappa\varrho _H^{5/3}~~;~~\kappa = p_{H_0} / \varrho_{H_0}^{5/3} ~.
    \end{aligned}
    \end{equation}
Note that the previous system is overdetermined. We compare the above Newtonian Friedmann equations with the Friedmann equations arising in general relativity with pressure:
\begin{equation}\label{ENJFRiedGR}
    \begin{aligned}
    3\frac{\ddot{a}}{a} + 4\pi G (\varepsilon_H + 3p_H) -\Lambda &= 0 ~;\\
3H^2+ \frac{c^2}{2}{\mathcal R}_H - 8\pi G\varepsilon_H - \Lambda &= 0~;\\
 \dot{\varepsilon}_H + 3H(\varepsilon_H +p_H)  = 0~;~ \varepsilon_H = \frac{1}{(Ca^2 - \kappa)^{3/2}} &~~;~~C \equiv \frac{\varepsilon_{H_0}+p_{H_0}}{\varepsilon_{H_0}^{5/3}}  ~;\\
\frac{1}{p_H}\dot{p}_H = \frac{5}{3} \frac{1}{\varepsilon_H} \dot{\varepsilon}_H ~~;~~p_H = \kappa\varepsilon_H^{5/3} &~~;~~\kappa = p_{H_0} / \varepsilon_{H_0}^{5/3} ~,
    \end{aligned}
    \end{equation}
with $6k / a^2(t) = c^2{\mathcal R}_H $ representing a \textit{constant curvature space}, with the scalar curvature ${\mathcal R}_H$, and $\varepsilon_H$ the \textit{energy density}, which is equal to $\varrho_H$ when there is no pressure. 
Note that for $\kappa \rightarrow 0$ we recover $\varepsilon_H \propto a^{-3}$ as it must; also the representation of $\varepsilon_H$ requires a minimal scale factor $a_{\rm min} = \sqrt{\kappa /C}$ at which $\varepsilon_H$ diverges (the kinetic pressure is stiffer than the matter and dominates at early times).
The equation of state for a homogeneous-isotropic kinetic pressure has been proven in \cite{EhlersRienstra}. With the help of this equation of state we have above integrated the conservation law for the energy density. We see that, in Newton's theory, the kinetic pressure is not a source of the gravitational field. For simplicity, and in conformity with the common use of a background in Newtonian cosmology, we will consider the background cosmology to be a solution of the Newtonian system in what follows.

\subsubsection{The Euler-Newton-Jeans system for deviations off the background}

To introduce the deviation fields, we need to define the \textit{comoving Eulerian coordinates} ${\bf q} : = {\bf x} / a(t)$ that are Lagrangian for the background cosmology \eqref{ENJFRiedGR}. In terms of these new coordinates we then split the dependent variables as follows: 
\begin{subequations}
\label{Split}
\begin{align}
    \varrho(\boldsymbol{q},t) &\equiv \varrho_H(t)[1+\delta(\boldsymbol{q},t)]~;\\
    \boldsymbol{v}(\boldsymbol{q},t) &\equiv \boldsymbol{v}_H + \boldsymbol{u}(\boldsymbol{q},t) ~~;~~ \boldsymbol{v}_H = \dot{a}\boldsymbol{q} ~;\\
    \boldsymbol{g}(\boldsymbol{q},t) &\equiv \boldsymbol{g}_H + \boldsymbol{w}(\boldsymbol{q},t)~~;~~ \boldsymbol{g}_H = \ddot{a}\boldsymbol{q} ~;\\
    \boldsymbol{\Pi} (\boldsymbol{q},t) &\equiv \boldsymbol{\pi}_H + \boldsymbol{\pi}(\boldsymbol{q},t) 
  ~~;~~\boldsymbol{\pi}_H \equiv p_H \boldsymbol{\delta} \ ,
\end{align}
where $\boldsymbol{\delta} = \delta_{ij}\, a^2\,\rd q_i \rd q_j = \delta_{ij}\, \rd x_i \rd x_j$ is the Euclidean $3-$metric, and
where $\pi_{ij} (\boldsymbol{q},t) = \varrho (\overline{u_i u_j} - {\bar u}_i {\bar u}_j) - p_H \delta_{ij} = \Pi_{ij} - p_H \delta_{ij}$ (since $\boldsymbol{v}-\boldsymbol{\bar v} =\boldsymbol{u}-\boldsymbol{\bar u}$), and $\boldsymbol{\nabla} \cdot \boldsymbol{\Pi} = (1/a) \boldsymbol{\nabla}_{\bf q} \cdot \boldsymbol{\pi}$.
\end{subequations}
The deviation fields $\delta$, ${\bf u}$ and ${\bf w}$ are commonly called density contrast, peculiar-velocity and peculiar-acceleration, respectively; we call $\boldsymbol{\pi}$ the reduced second moment peculiar-tensor; the derivative operators transform as follows:
\begin{subequations}
\label{comovingnabla}
\begin{align}
\boldsymbol{\nabla}_{\boldsymbol{q}} &= a(t)\boldsymbol{\nabla} ~;\\
\frac{\partial}{\partial t}\Big\vert_{\boldsymbol{q}} &= \frac{\rd}{\rd t}\Big\vert_{\boldsymbol{v}_H} = \frac{\partial}{\partial t}\Big\vert_{\boldsymbol{x}} + (\boldsymbol{v}_H\cdot\boldsymbol{\nabla}) \boldsymbol{v}_H ~;\\
\textrm{or}~~\frac{\partial}{\partial t}\Big\vert_{\boldsymbol{x}} &= \frac{\partial}{\partial t}\Big\vert_{\boldsymbol{q}} - H(t)\,\boldsymbol{q}\cdot\boldsymbol{\nabla}_{\boldsymbol{q}} ~. 
\end{align}
We notice that the Lagrangian derivative has the following invariant property with respect to this change:
\begin{equation}
\label{invariancederivative}
\frac{\mathrm d}{{\mathrm d}t}\Big\vert_{\bf v} = \frac{\mathrm d}{{\mathrm d}t}\Big\vert_{{\bf u}/a} \equiv \frac{\mathrm d}{{\mathrm d}t} \ ,
\end{equation}
where the last equation abbreviates the Lagrangian derivative in both coordinate frames.
\end{subequations}
We can derive the following system for the deviation fields by splitting the Euler-Newton-Jeans system (\ref{ENJ1}) (for vanishing reduced third moment) via \eqref{Split}, \eqref{comovingnabla}, and by inserting the background part, governed by the Friedmann equations \eqref{ENJFRiedGR}, to obtain the Euler-Newton-Jeans system for the peculiar-dynamics (dropping the divergence of the third reduced velocity moment):
\begin{equation}\label{ENJcom1}
    \begin{aligned}
 \frac{\rd}{\rd t}\delta &= - \frac{1}{a}(1+\delta)\boldsymbol{\nabla}_{\boldsymbol{q}}\cdot\boldsymbol{u}~;\\
 \frac{\mathrm{d}}{\mathrm{d}t} \boldsymbol{u} +H \boldsymbol{u} &= \boldsymbol{w} - \frac{1}{a}\frac{1}{\varrho_H (1+\delta)} \boldsymbol{\nabla}_{\bf q}\cdot \boldsymbol{\pi}~;\\ 
\frac{\mathrm{d}}{\mathrm{d}t} \pi_{ij} + 5H \pi_{ij}= &- \frac{1}{a} \Big[\frac{\partial u_i}{\partial q_k}\pi_{jk} + \frac{\partial u_j}{\partial q_k}\pi_{ki} + \frac{\partial u_k}{\partial q_k}\pi_{ij}\Big] \\ 
& - \frac{p_H}{a} \Big[ \frac{\partial u_k}{\partial q_k}\,\delta_{ij} + \frac{\partial u_i}{\partial q_j} + \frac{\partial u_j}{\partial q_i} \Big]~; \\
 \boldsymbol{\nabla}_{\boldsymbol{q}}\times \boldsymbol{w} &= {\bf 0} ~;\\
 \boldsymbol{\nabla}_{\boldsymbol{q}}\cdot\boldsymbol{w} &= -4\pi G a\, \varrho_H {\delta} ~.
    \end{aligned}
    \end{equation}
The homogeneous contribution to the reduced second moment tensor vanishes in the source of the Euler-Jeans equation due to the spatial derivative; the second line in the evolution equation for this tensor describes the coupling of the background dispersion to the peculiar-expansion and peculiar-shear, and this line would vanish for zero background dispersion.

 Again, the special case of an isotropic distribution of the peculiar-dispersion tensor on an isotropic background (\ref{anisPI}), with stress tensor having a state equation $p = \beta(\varrho)$, the Euler-Newton-Jeans system (\ref{ENJcom1}) becomes a system for $\boldsymbol{u}(\boldsymbol{q},t) $, $\boldsymbol{w}(\boldsymbol{q},t)$ and ${\delta}(\boldsymbol{q},t)$:
\begin{equation}\label{ENJcom2}
    \begin{aligned}
 \frac{\rd}{\rd t}\delta &= - \frac{1}{a}(1+\delta)\boldsymbol{\nabla}_{\boldsymbol{q}}\cdot\boldsymbol{u}~;\\
 \frac{\mathrm{d}}{\mathrm{d}t} \boldsymbol{u} +H \boldsymbol{u} &= \boldsymbol{w} - \frac{\beta'(\varrho)}{a\varrho}\boldsymbol{\nabla}_{\boldsymbol{q}}\;\varrho~;\\ 
 \boldsymbol{\nabla}_{\boldsymbol{q}}\times \boldsymbol{w} &= {\bf 0} ~;\\
 \boldsymbol{\nabla}_{\boldsymbol{q}}\cdot\boldsymbol{w} &= -4\pi G a\, \varrho_H {\delta} ~.
    \end{aligned}
    \end{equation}
which is the Euler-Newton-Jeans system for the peculiar-dynamics. We can also compute the Poisson equation for a peculiar-gravitational potential $\phi$, $\boldsymbol{w} =: - a^{-1}\boldsymbol{\nabla}_{\bf q}\, \phi$:
\begin{equation}\label{Poissoncom1}
    \Delta_{\bf q} \phi = a^2(t)4\pi G\varrho_H(t) \delta(\boldsymbol{q},t) ~.
\end{equation}
We now express the principal scalar invariants of the velocity gradient in terms of those of the peculiar-velocity gradient (invariants always refer to the tensor field that follows in brackets):
\begin{subequations}\label{invcom2}
\begin{align}
    \inI(\boldsymbol{v}) &= 3H(t) + \inI(\tfrac{\boldsymbol{u}}{a})  ~;\\
    \inII(\boldsymbol{v}) &= 3H^2(t)+2H(t)\inI(\tfrac{\boldsymbol{u}}{a})+\inII(\tfrac{\boldsymbol{u}}{a})~.
\end{align}
\end{subequations}
We remark that the local quantity $\CQ^{\rm loc}$ is invariant with respect to the change of $\bf v$ to $\bf u$:
\begin{equation}\label{Qinvcom}
    \CQ^{\rm loc}  = 2 \inII(\boldsymbol{v}) - \frac{2}{3}\inI^2(\boldsymbol{v}) = 2 \inII(\tfrac{\boldsymbol{u}}{a}) - \frac{2}{3}\inI^2(\tfrac{\boldsymbol{u}}{a})~.
\end{equation}
This local quantity, when averaged, should not be confused with the so-called \textit{kinematical backreaction} $\mathcal{Q}_{\mathcal{D}}$ introduced in the next subsection.

\subsubsection{Averaged system}

First, we need to define the averaging operation. For a scalar field $\chi(x,t)$ over a volume $V_\CD(t)$ we consider the spatial volume average, first in the non-comoving Eulerian coordinates:
\begin{align}
\langle\chi(x,t)\rangle_\CD = \frac{1}{V_\CD(t)}\int_{\CD_t} \chi(x,t)\mathrm{d}^3x &\\
 ~~\mathrm{ with }~~ V_\CD(t) =\int_{\CD_t} \mathrm{d}^3x = \int_{\CD_i} J \mathrm{d}^3X ~; \\
a_\CD(t) = \Big(\frac{V_\CD(t)}{V_{\CD_i}} \Big)^{\frac{1}{3}}&~; \\
\langle\theta\rangle_\CD = 3\frac{{\dot a}_\CD(t)}{a_\CD(t)}&
~, \end{align}
where we introduced a volume scale factor, $a_\CD$ and the Jacobian of the transformation from Eulerian to Lagrangian coordinates, $J \equiv \det ( f_{i\vert k})$, where the Eulerian coordinates are considered as vector fields of Lagrangian coordinates (trajectories), $x_i = f_i (X_k ,t)$, and where the vertical slash denotes derivative with respect to Lagrangian coordinates.
The averaging operation and the time-evolution do not commute \cite{BuchertEhlers97}:
\begin{align}
\label{eq:commutation}
\frac{\mathrm{d}}{\mathrm{d}t} 
\langle\chi(x,t)\rangle_{\CD} = \frac{\mathrm{d}}{\mathrm{d}t}\Big(\frac{1}{V_\CD(t)}\int_{D_t} \chi(x,t)\mathrm{d}^3x \Big)\qquad\qquad\quad\nonumber \\
= - \frac{\dot{V}_\CD}{V_\CD(t)}\frac{1}{V_\CD(t)}\Big(\int_{D_t} \chi(x,t)\mathrm{d}^3x \Big) + \frac{1}{V_\CD(t)}\int_{\CD_i} \frac{\mathrm{d}}{\mathrm{d}t}(\chi(x,t)J)\mathrm{d}^3X  \nonumber \\
\Rightarrow \  \frac{\mathrm{d}}{\mathrm{d}t}\langle\chi\rangle_\CD - \langle\frac{\mathrm{d}}{\mathrm{d}t}\chi\rangle_\CD = \langle\chi\theta\rangle_\CD - \langle\theta\rangle_\CD\langle\chi\rangle_\CD
~.
\end{align}
For the averaged continuity equation, setting $\chi = \varrho$, we obtain:
\begin{equation}
\label{avcontinuity}
\frac{\mathrm{d}}{\mathrm{d}t}\langle\varrho\rangle_\CD + \langle\theta\rangle_\CD\langle\varrho\rangle_\CD =  \Big\langle\frac{\mathrm{d}}{\mathrm{d}t}\varrho + \varrho \theta\Big\rangle_\CD = 0
~. \end{equation}
From \eqref{avcontinuity}, $\langle\varrho\rangle_\CD = M_\CD / (V_i a_\CD(t)^3)$, with $M_\CD$ the total mass enclosed within the domain $\CD$.
Setting $\chi = \theta$, we first obtain:
\begin{equation}\label{raychaudglob1}
  \frac{\mathrm{d}}{\mathrm{d}t}\langle\theta\rangle_\CD - \langle\frac{\mathrm{d}}{\mathrm{d}t}\theta\rangle_\CD = \langle\theta^2\rangle_\CD - \langle\theta\rangle^2_\CD  ~.
\end{equation}
Inserting the local Raychaudhuri equation (\ref{Raychaudloc}) the evolution equation for the averaged expansion rate reads:
\begin{equation}\label{raychaudglob2}
\frac{\mathrm{d}}{\mathrm{d}t}\langle\theta\rangle_\CD = \Lambda -4\pi G\langle\varrho\rangle_\CD -\frac{1}{3}\langle\theta\rangle^2_\CD + \CQ_\CD~.
\end{equation}
With 
\begin{equation}
\frac{\mathrm{d}}{\mathrm{d}t}\langle\theta\rangle_\CD = 3\frac{{\ddot a}_\CD(t)}{a_\CD(t)} - \frac{1}{3}\langle\theta\rangle_\CD^2
~, \end{equation}
the averaged Raychaudhuri equation assumes the form of a generalization of the Friedmannian acceleration law:
\begin{equation}
\label{avacceleration}
3\frac{{\ddot a}_\CD(t)}{a_\CD(t)} - \Lambda +4\pi G \langle\varrho\rangle_\CD = \CQ_\CD
~, \end{equation}
The term $\CQ_\CD$ is the \textit{kinematical backreaction} encoding the deviations from the Friedmannian law,  
\begin{equation}
\label{Q}
\CQ_\CD = \frac{2}{3}\Big[\langle(\theta - \langle\theta\rangle_\CD)^2\rangle_\CD\Big] + 2\langle\omega^2-\sigma^2\rangle_\CD - \Big\langle \frac{\partial }{\partial x_i}\Big( \frac{1}{\varrho}\Psi_i \Big)\Big\rangle_\CD
~. \end{equation}
Multiplying by $2{\dot a}_\CD a_\CD$, we can formally integrate Equation~(\ref{avacceleration}) to obtain a generalization of the Friedmannian expansion law:
\begin{equation}
\label{avexpansion}
3\frac{{\dot a}_\CD^2+k_\CD}{a_\CD^2} - 8\pi G\langle\varrho\rangle_\CD - \Lambda = \frac{1}{a_\CD^2}\int_{t_i}^t\mathrm{d}t' \CQ_\CD\frac{\mathrm{d}}{\mathrm{d}t'}a_\CD^2(t')
~, \end{equation}
with $k_\CD$ a domain-dependent integration constant. 
We divide the above equation by $H_\CD^2$, with $H_\CD \equiv\dot{a}_\CD / a_\CD$ to define dimensionless average characteristics that correspond to `cosmological parameters':
\begin{equation}\label{avexpansionOmega}
   \Omega_k^\CD + \Omega_m^\CD + \Omega_{\Lambda}^\CD + \Omega_Q^\CD =1 \ ,
\end{equation}
with the definitions: $\Omega_m^\CD \equiv 8\pi G\langle\varrho\rangle_\CD / 3H_\CD^2$ for matter, $\Omega_k^\CD \equiv - k_\CD / (a_\CD^2H_\CD^2)$ for a scale-dependent constant curvature, $\Omega^\CD_{\Lambda} \equiv \Lambda / 3H_\CD^2$, and the additional backreaction functional:
$$\Omega_\CQ^\CD \equiv \frac{1}{3H_\CD^2a_\CD^2}\int_{t_i}^t\mathrm{d}t' \CQ_\CD\frac{\mathrm{d}}{\mathrm{d}t'}a_\CD^2(t')\ .$$ 

\noindent These dimensionless quantities allow us to draw conclusions about the energy content in the averaged Universe. We also see that not only the fluctuations are spatially added up to the given averaging scale, but also the time-history of the fluctuations influences the effective expansion law. 
The set of equations \eqref{avcontinuity}, \eqref{avacceleration}, \eqref{avexpansion} together with the \textit{kinematical backreaction} term \eqref{Q} generalize the averaged equations to include velocity dispersion, valid for general collisionless Newtonian self-gravitating systems \cite{BuchertEhlers97}, \cite{BKS}, \cite{Buchert18}. In Appendix \ref{appB} we recall known theorems about the backreaction term $\CQ_\CD$, extended for the case with velocity dispersion. 

\section{Average properties of volume-partitioned Newtonian cosmologies}
\label{s-partitioning}

The volume partitioning approach to Newtonian cosmology is based on dividing space into sub-domains that are in principle easier to model than the global properties of the whole Universe, which can then be inferred with the use of algebraic summation rules. We will discuss `separate Universe conjectures' that can be concisely formulated within exact and dynamical volume partitionings. The partitioning rules were derived within general relativity, first proposed for a two-scale partitioning by \cite{W07} and follow-up papers (see, e.g. \cite{W14} and references therein) including the backreaction term between the two domains, and later developed for the general case for hypersurfaces including also backreaction on the individual domains in \citet{buchertcarfora}, then for a fully dynamical multi-scale partitioning in (\cite{multiscale}). As the Newtonian average equations are similar in form to the relativistic ones, the partitioning relations carry over to the Newtonian framework. The Newtonian partitioning formulas are explicitly given in the next subsection.

\subsection{Generic volume partitioning of Newtonian cosmologies}

We will first present a general multi-scale partitioning before restricting the formalism to the two-scale problem of subregion unions that may be characterized by typical \textit{underdense} and \textit{overdense} unions of spatial domains. We will translate the investigations in \cite{buchertcarfora} and \cite{multiscale}
into the Newtonian framework.

\subsubsection{Multi-scale partitioning method}

The idea of general volume partitioning of cosmological hypersurfaces is to divide a compact domain $\mathcal D$,
covering a homogeneity-scale domain or a whole spatial section of a compact universe model (then denoted by $\Sigma \equiv \CD$) into $\ell$ subregions ${\mathcal F}_\ell$. We also assume that the subdomain boundaries never intersect, i.e. at all times we assume
that the intersection between two subdomains ${\mathcal F}_\ell \cap {\mathcal F}_m$ is a piecewise smooth surface with volume measure zero. 
With this regularity condition, we have the volume additivity rule\footnote{This is true for domains with Lagrangian boundaries, since there will be no overlap of the domains for regular solutions. Shell-crossing may, however, lead to overlap of the domains, and the volume additivity has to be replaced by Minkowski additivity \cite{Minkowski}, implying the task to deal with the evolution of the intersecting part. We will here confine ourselves to regular solutions for which this assumption is valid in general.} :
\begin{equation}\label{volpart1}
V_{\mathcal D} = \sum_\ell V_{{\mathcal F}_\ell} ~,
\end{equation}
with the volume being $V_{{\mathcal F}_{\ell}} := \int_{{\mathcal F}_{\ell}} \rd ^3 x$. For every scalar function $\chi$ on every domain ${\mathcal F}_{\ell}$ we then have:
\begin{equation}\label{volpart2}
\langle \chi \rangle _{\CD} = \sum_{\ell} \lambda_{{\mathcal F}_{\ell}} \langle \chi \rangle_{{\mathcal F}_{\ell}} ~,
\end{equation}
where we defined the volume fractions:
\begin{equation}\label{volpart3}
\lambda_{{\mathcal F}_{\ell}} \equiv \frac{V_{{\mathcal F}_{\ell}}}{V_{\CD}}  ~.
\end{equation}
We can also define a volume scale factor $a_{{\mathcal F}_{\ell}}$ for every domain ${\mathcal F}_{\ell}$ . In order to use the additivity of the volumes, we will define them by dividing every volume by the global one at the initial time\footnote{This definition has the obvious advantage that, although we generally apply the formalism to expanding universe models, a subdomain may have vanishing Hubble rate. (Exception would be a globally static model that we exclude here.)}:
\begin{equation}\label{volpart4}
    a_{{\mathcal F}_{\ell}} := \Big(\frac{V_{{\mathcal F}_{\ell}}}{V_{\CD_i}}\Big)^{\frac{1}{3}} ~.
\end{equation}
Using the volume additivity we have:
\begin{equation}\label{volpart5}
    a_{\CD}^3 = \sum_{\ell} a^3_{{\mathcal F}_{\ell}} ~.
\end{equation}
We also define the effective Hubble ratio in the subregions:
\begin{equation}\label{volpart6}
    H_{{\mathcal F}_{\ell}} = \frac{\dot a_{{\mathcal F}_{\ell}}}{a_{{\mathcal F}_{\ell}}} = \frac{1}{3}\langle \theta \rangle_{{\mathcal F}_{\ell}}\ .
\end{equation}
We then look at Equation (\ref{Q})
and apply Equation (\ref{volpart2}) to the two last terms, $\langle\omega^2-\sigma^2\rangle_\CD$ and $\langle \frac{\partial }{\partial x_i}\Big( \frac{1}{\varrho}\Psi_i \Big)\rangle_\CD$ which behave linearly with respect to the volume partitioning. The first term contains a nonlinear part, $\langle \theta\rangle_{\CD}^2$, which does not decompose linearly in the partitioned volume and deduce that $\CQ_\CD \neq \sum_{\ell} \lambda_{{\mathcal F}_{\ell}} \CQ_{{\mathcal F}_{\ell}}$, contrary to the preliminary expectation in Equation (61) of \cite{BKS}. Calculating the nonlinear part we obtain:
\begin{equation}\label{volpart7}
    \CQ_{\CD} = \sum_{\ell} \lambda_{{\mathcal F}_{\ell}} \CQ_{{\mathcal F}_{\ell}} + 6\sum_{\ell} \lambda_{{\mathcal F}_{\ell}} H^2_{{\mathcal F}_{\ell}} - 6 H_{\CD}^2 ~,
\end{equation}
where $\CQ_{{\mathcal F}_{\ell}} = \frac{2}{3}\Big[\langle(\theta - \langle\theta\rangle_{{\mathcal F}_{\ell}})^2\rangle_{{\mathcal F}_{\ell}}\Big] + 2\langle\omega^2-\sigma^2\rangle_{{\mathcal F}_{\ell}} - \langle \frac{\partial }{\partial x_i}\Big( \frac{1}{\varrho}\Psi_i \Big)\rangle_{{\mathcal F}_{\ell}}$, and $H_{\CD} = \sum_{\ell} \lambda_{{\mathcal F}_{\ell}} H_{{\mathcal F}_{\ell}}$.

Using $\sum_\ell\lambda_{{\mathcal F}_\ell}H^{2}_{{\mathcal F}_\ell}-H^{2}_\CD
=\tfrac12\sum_{\ell\neq m}\lambda_{{\mathcal F}_\ell}\lambda_{{\mathcal F}_m}(H_{{\mathcal F}_\ell}-H_{{\mathcal F}_m})^{2}$, Equation~\eqref{volpart7} can be rewritten as:
\begin{equation}\label{volpart8}
    \CQ_{\CD} = \sum_{\ell} \lambda_{{\mathcal F}_{\ell}} \CQ_{{\mathcal F}_{\ell}} + 3\sum_{l\neq m} \lambda_{{\mathcal F}_{\ell}}\lambda_m(H_{{\mathcal F}_{\ell}} - H_m )^2 ~.
\end{equation}
Differentiating the volume additivity  (\ref{volpart5}) with respect to time and using the results above, we obtain a relation between global and subdomain effective (volume scale factor) accelerations:
\begin{equation}\label{volpart9}
    \frac{\ddot{a}_\CD}{a_\CD} = \sum_{\ell} \lambda_{{\mathcal F}_{\ell}} \frac{\ddot{a}_{{\mathcal F}_{\ell}}}{a_{{\mathcal F}_{\ell}}} + 2\sum_{\ell}\lambda_{{\mathcal F}_{\ell}} H^2_{{\mathcal F}_{\ell}} -2H_\CD^2 ~.
\end{equation}
As a consequence we infer that the global effective acceleration is not simply due to the sum of the regional ones, but it also depends on the relative rate of expansions between regions of different type. Even with deceleration on average on all the subregions, the second term might lead to a global effective acceleration, or to global stationarity for example. We have the following system of \textit{splitting equations}:
\begin{equation}\label{volpartsys1}
    \begin{aligned}
\langle \varrho \rangle_\CD &= \sum_{\ell} \lambda_{{\mathcal F}_{\ell}} \langle \varrho \rangle_{{\mathcal F}_{\ell}}~; \\
H_\CD &= \sum_{\ell} \lambda_{{\mathcal F}_{\ell}} H_{{\mathcal F}_{\ell}} ~; \\
\frac{\ddot{a}_\CD}{a_\CD} &= \sum_{\ell} \lambda_{{\mathcal F}_{\ell}} \frac{\ddot{a}_{{\mathcal F}_{\ell}}}{a_{{\mathcal F}_{\ell}}} + 2\sum_{\ell}\lambda_{{\mathcal F}_{\ell}} H^2_{{\mathcal F}_{\ell}} -2H_\CD^2 ~;\\
\CQ_\CD &= \sum_{\ell} \lambda_{{\mathcal F}_{\ell}} \CQ_{{\mathcal F}_{\ell}} + 6\sum_{\ell} \lambda_{{\mathcal F}_{\ell}} H^2_{{\mathcal F}_{\ell}} - 6 H_{\CD}^2 ~.    
\end{aligned}
\end{equation}
Since every subregion union satisfies the averaged Raychaudhuri and averaged continuity equations, we obtain the corresponding equations on the global domain. We can also derive an evolution equation for the volume fraction $\lambda_{{\mathcal F}_{\ell}}$: with
\begin{equation}\label{volpart11}
    H_{\CD} = \frac{\dot{a}_\CD}{a_\CD} = \sum_{\ell} \lambda_{{\mathcal F}_{\ell}} \frac{\dot{a}_{{\mathcal F}_{\ell}}}{a_{{\mathcal F}_{\ell}}} = \sum_{\ell} \lambda_{{\mathcal F}_{\ell}} H_{{\mathcal F}_{\ell}} ~,
\end{equation}
and performing the time-derivative we obtain:
\begin{align}
\frac{\ddot{a}_{\CD}}{a_\CD} - \Big(\frac{\dot{a}_\CD}{a_\CD}\Big)^2 &= \sum_{\ell} \Big [ \lambda_{{\mathcal F}_{\ell}} \frac{\ddot{a}_{{\mathcal F}_{\ell}}}{a_{{\mathcal F}_{\ell}}} - \lambda_{{\mathcal F}_{\ell}}\Big(\frac{\dot{a}_{{\mathcal F}_{\ell}}}{a_{{\mathcal F}_{\ell}}}\Big)^2 +\dot{\lambda}_{{\mathcal F}_{\ell}} \frac{\dot{a}_{{\mathcal F}_{\ell}}}{a_{{\mathcal F}_{\ell}}} \Big] ~; \nonumber\\
\label{volpart13} \frac{\ddot{a}_{\CD}}{a_\CD} &= \sum_{\ell} \lambda_{{\mathcal F}_{\ell}} \frac{\ddot{a}_{{\mathcal F}_{\ell}}}{a_{{\mathcal F}_{\ell}}} - \sum_{\ell} \lambda_{{\mathcal F}_{\ell}} H^2_{{\mathcal F}_{\ell}} + \sum_{\ell}\dot{\lambda}_{{\mathcal F}_{\ell}} H_{{\mathcal F}_{\ell}}+H^2_\CD ~.
\end{align}
By comparison of Equations (\ref{volpart9}) and (\ref{volpart13}) we find an equation for the evolution of $\lambda_{{\mathcal F}_{\ell}}$:
\begin{equation}\label{volpart14}
    \sum_{\ell}\dot{\lambda}_{{\mathcal F}_{\ell}} H_{{\mathcal F}_{\ell}} = 3 \sum_{\ell} \lambda_{{\mathcal F}_{\ell}} H_{{\mathcal F}_{\ell}}^2 - 3H^2_{\CD} ~.
\end{equation}
We also introduce the four averaged characteristics in the partitioned volume as in Equation (\ref{avexpansionOmega}). The volume partitioned characteristics are defined as:
\begin{eqnarray}\label{avexpansionOmegaPart}
    &\Omega_m^{{\mathcal F}_{\ell}} := \frac{8\pi G\langle\varrho\rangle_{{\mathcal F}_{\ell}}}{3H^2_{\CD}} ~~;~~ \Omega_\Lambda^{{\mathcal F}_{\ell}} := \frac{\Lambda}{3H^2_{\CD}} ~~;~~ \Omega_k^{{\mathcal F}_{\ell}} := -\frac{k_{{\mathcal F}_{\ell}}}{a^2_{{\mathcal F}_{\ell}}H^2_{\CD}} ~~;~~ \nonumber\\
   &\Omega_Q^{{\mathcal F}_{\ell}} := \frac{1}{3H^2_{\CD}a^2_{{\mathcal F}_{\ell}}}\int_{t_i}^t\mathrm{d}t' \CQ_{{\mathcal F}_{\ell}}\frac{\mathrm{d}}{\mathrm{d}t'}a_{{\mathcal F}_{\ell}}^2(t')\ .
\end{eqnarray}
Their sum differs from Equation~(\ref{avexpansionOmega}) due to our definition (\ref{volpart4}):
\begin{equation}\label{avexpansionOmegaPart2}
     \Omega_m^{{\mathcal F}_{\ell}} + \Omega_\Lambda^{{\mathcal F}_{\ell}} + \Omega_k^{{\mathcal F}_{\ell}} + \Omega_Q^{{\mathcal F}_{\ell}} = \frac{H^2_{{\mathcal F}_{\ell}}}{H^2_\CD} \ .
\end{equation}
This constraint agrees with (\ref{avexpansionOmega}) in the case where there is no partitioning, ${\mathcal F}_{\ell} = \CD$. 

\subsubsection{Two-scale partitioning of a global spatial hypersurface}

We developed a formalism for a partitioning with $\ell$ subregions ${\mathcal F}_{\ell}$. We are now
ready to use the partitioning method on two different main subregion types, called \textit{overdensity} and \textit{underdensity}. To be more precise we define these regions at an initial time as $M$ being the union of the regions with initial overdensity, and $E$ its complement in relation to the global domain being the union of all underdense regions. We henceforth identify the domain $\CD$ with the whole spatial hypersurface $\Sigma$ in the splitting equations and in the calculations before. 

The following formalism will describe the evolution of the union of all underdense or overdense regions $E$ or $M$, respectively, \textit{on average}. Considering a partitioning into such `typical' domains entails an
implicit cosmological principle on regional scales: with a `typical region' we average over individual $M$- or $E$- regions. Observations show that, at present time, we live in a void-dominated region, so that we expect the volume fraction of $M$ regions to decrease in the course of structure formation. Taking into account the fact that $M$ is the complement of $E$ at every time, we have $\lambda_M = 1 - \lambda_E$. With these assumptions, our partitioning formulas simplify, and the system of splitting equations can be written:
\begin{equation}\label{volpartsys2}
    \begin{aligned}
\langle \varrho \rangle_\Sigma &=  \lambda_E \langle \varrho \rangle_{E} + (1-\lambda_E) \langle \varrho \rangle_{M}~; \\
H_\Sigma &=  \lambda_E H_{E} + (1-\lambda_E) H_{M}~; \\
\frac{\ddot{a}_\Sigma}{a_\Sigma} &=  \lambda_E \frac{\ddot{a}_E}{a_E} + (1-\lambda_E) \frac{\ddot{a}_M}{a_M} + 2 \lambda_E(1-\lambda_E)(H_E-H_M)^2 ~;\\
\CQ_\Sigma &=  \lambda_E \CQ_E + (1-\lambda_E) \CQ_{M} + 6 \lambda_E(1-\lambda_E)(H_E-H_M)^2  ~.
\end{aligned}
\end{equation}
Equation (\ref{volpart14}) for the evolution of the volume fraction now reads:
\begin{equation}\label{volpart21}
    \dot{\lambda}_E = 3\lambda_E(1-\lambda_E)(H_E-H_M) ~.
\end{equation}
All the equations above are exact and do not involve approximations or restrictions for regular solutions.

\subsubsection{Hubble-comoving formulation}
\label{Hubblecomoving}

Hereon we define the averaging operation in comoving coordinates. We first define the volume in comoving coordinates:
\begin{equation}\label{AverageComoving1}
    V_{\CD_q} \equiv \int_{\CD_q}\rd^3 q ~,
\end{equation}
where the integral is now over a domain that is point-wise rescaled by $a(t)$, $\CD_q \equiv \CD / a(t)$. We then define the averaging operation of a scalar field $\chi$:
\begin{equation}\label{AverageComoving2}
    \langle \chi \rangle_{\CD_q} \equiv \frac{1}{V_{\CD_q}}\int_{\CD_q} \chi \,\rd^3 q ~,
\end{equation}
Because of a merely time-dependent rescaling, the averaging operation in Hubble-comoving coordinates is equivalent to averaging in non-comoving coordinates,
\begin{equation}\label{AverageComoving3}
    \langle \chi \rangle_{\CD_q} = \frac{1}{a^3V_{\CD_q}}\int_{\CD_q}a^3\chi\,\rd^3q = \frac{1}{V_\CD}\int_{\CD}\chi \,\rd^3x = \langle \chi \rangle_\CD ~.
\end{equation}
A straightforward calculation shows that the kinematical backreaction term (\ref{Q}) cancels the Hubble terms and is also invariant (\textit{cf.} (\ref{invcom3})):
\begin{align}\label{invcom32}
2\langle\inII(\bm v)\rangle_\CD=6H^{2}+4H\langle\inI(\tfrac{\bm u}{a})\rangle_\CD+2\langle\inII(\tfrac{\bm u}{a})\rangle_\CD \; ,\nonumber\\
\tfrac23\langle\inI(\bm v)\rangle_\CD^{2}=6H^{2}+4H\langle\inI(\tfrac{\bm u}{a})\rangle_\CD+\tfrac23\langle\inI(\tfrac{\bm u}{a})\rangle_\CD^{2} \; ,\nonumber\\
 \Rightarrow \   \CQ_\CD = 2\langle\inII(\tfrac{\boldsymbol{u}}{a})\rangle_\CD -\frac{2}{3}\langle\inI(\tfrac{\boldsymbol{u}}{a})\rangle^2_\CD- \frac{1}{a}\Big\langle \frac{\partial }{\partial q_i}\Big( \frac{1}{\varrho}\Psi_i\Big)\Big\rangle_\CD ~.
\end{align} 
In the following we will use the previous notation $\langle \chi \rangle_\CD$ whenever we consider spatial averages also in the Hubble-comoving framework. Recall in this context Equation (\ref{invariancederivative}) for the invariance of the Lagrangian time-derivative. 

For any given domain, e.g. ${\mathcal F}_{\ell}$, we can split the Hubble function as:
\begin{equation}\label{ComovingHubble}
    H_{{\mathcal F}_{\ell}} = H(t) + \mathfrak{h}_{{\mathcal F}_{\ell}} (t) ~,
    \end{equation}
where we introduced 
a peculiar-Hubble rate via a rescaled volume scale factor $\mathfrak{a}_{{\mathcal F}_{\ell}} (t) := 
a_{{\mathcal F}_{\ell}} (t) / a (t)$, 
\begin{equation}
\mathfrak{h}_{{\mathcal F}_{\ell}} (t) : = \frac{\dot{\mathfrak{a}}_{{\mathcal F}_{\ell}}}{\mathfrak{a}_{{\mathcal F}_{\ell}}} = H_{{\mathcal F}_{\ell}}  - H ~,
\end{equation}
and a peculiar-acceleration functional, 
\begin{equation}
\mathcal{A}_{{\mathcal F}_{\ell}} (t): = \frac{{\ddot a}_{{\mathcal F}_{\ell}}}{a_{{\mathcal F}_{\ell}}} - \frac{\ddot a}{a} \ne \frac{\ddot{\mathfrak{a}}_{{\mathcal F}_{\ell}}}{\mathfrak{a}_{{\mathcal F}_{\ell}}}~,
\end{equation}
satisfying the identities,
\begin{equation}
\mathcal{A}_{{\mathcal F}_{\ell}} = \dot{\mathfrak{h}}_{{\mathcal F}_{\ell}} + \mathfrak{h}_{{\mathcal F}_{\ell}}^2 + 2H \mathfrak{h}_{{\mathcal F}_{\ell}} = \frac{1}{a^2 \mathfrak{a}_{{\mathcal F}_{\ell}}} \frac{d}{dt}\left( a^2 \mathfrak{a}_{{\mathcal F}_{\ell}} \mathfrak{h}_{{\mathcal F}_{\ell}}\right)~. 
\end{equation}
The system of splitting equations restricted to the two-scale problem, (\ref{volpartsys2}), becomes:
\begin{equation}\label{volpartsyspec}    
    \begin{aligned}
\langle \delta \rangle_{\Sigma} &=  \lambda_E \langle \delta \rangle_{E} + (1-\lambda_E) \langle \delta \rangle_{M}\quad ; \quad 
\mathfrak{h}_{\Sigma} =  \lambda_E \mathfrak{h}_{E} + (1-\lambda_E) \mathfrak{h}_{M}~; \\
\mathcal{A}_\Sigma &=  \lambda_E \mathcal{A}_E + (1-\lambda_E) \mathcal{A}_M + 2 \lambda_E(1-\lambda_{E})(\mathfrak{h}_{E}-\mathfrak{h}_{M})^2 ~;\\
\CQ_{\Sigma} &=  \lambda_E \CQ_{E}+ (1-\lambda_E) \CQ_{M} + 6 \lambda_E(1-\lambda_E)(\mathfrak{h}_{E}-\mathfrak{h}_{M})^2  ~.
\end{aligned}
\end{equation}

\subsection{Closure conditions}
\label{s-closure}

Having developed an exact volume partitioning for regular dynamics of spatial domains and for general self-gravitating systems in Newtonian cosmology, we arrived at a system of a non-closed family of splitting equations. We will now consider the question of closing this family of equations through assumptions on regional domains. We first count the number of needed closure conditions. 

\subsubsection{The closure problem}

For the deviation field averages $\langle\delta\rangle_\CD$, $\mathfrak{h}_\CD$, $\mathcal{A}_\CD$ and $\CQ_\CD$ in Hubble-comoving coordinates, the system of equations \eqref{avcontinuity}, \eqref{avacceleration} with \eqref{Q}, 
and \eqref{avexpansion}, valid for any domain, becomes:
\begin{equation}\label{buchertequationspec}
    \begin{aligned}
\frac{\mathrm{d}}{\mathrm{d}t}\langle\delta\rangle_{\CD} + 3\mathfrak{h}_{\CD}(1+\langle\delta\rangle_{\CD}) &= 0 ~; \\
3\mathcal{A}_\CD +4\pi G \varrho_H \langle\delta\rangle_{\CD} &= \CQ_{\CD} ~; \\
\CQ_{\CD} = 2\langle\inII(\tfrac{\boldsymbol{u}}{a})\rangle_{\CD} -\frac{2}{3}\langle\inI(\tfrac{\boldsymbol{u}}{a})\rangle^2_{\CD} &- \frac{1}{a}\Big\langle \frac{\partial }{\partial q_i}\Big( \frac{1}{\varrho}\Psi_i\Big)\Big\rangle_{\CD} ~.
\end{aligned}
\end{equation}
Since $\mathfrak{h}_\CD$ and $\mathcal{A}_\CD$ can be constructed from the scaled volume scale factor $\mathfrak{a}_\CD$, we have three variables in the system \eqref{buchertequationspec} ($\mathfrak{a}_{\CD}$, $\CQ_{\CD}$, $\langle\delta\rangle_{\CD}$), but only two independent equations that describe the dynamics on any domain 
$\CD$: the averaged Raychaudhuri equation and the averaged continuity equation. This reasoning can be done on $\Sigma$, the global hypersurface, $E$- and $M$-regions, leading to $9$ dependent variables and $6$ independent equations for the three-domain problem. However, the split introduces a new variable, $\lambda_{E}$, for which we have one further equation \eqref{volpart21}, but which is not independent of the other equations. Therefore, \textit{a priori} we would need \textit{four} closure conditions for the three-domain problem.\\

\noindent{\bf Remark:} Since the second equation of the set~(\ref{buchertequationspec})
is just the difference between the exact regional acceleration law~\eqref{avacceleration} and the background acceleration law, one could drop the split into background and peculiar-fields entirely and work with \eqref{avacceleration} and the definitions $\langle\varrho\rangle_\CD = : \varrho_H (1 + \langle\delta\rangle_\CD)$ and $\CQ_\CD$ as defined in the set (\ref{buchertequationspec}), so that the two scalars $\langle\delta\rangle_\CD$ and $\CQ_\CD$ carry all peculiar-information.

\subsubsection{One closure condition needed in Newtonian cosmology}

Several closure conditions can be obtained using the theorems listed in Appendix \ref{appB}. Indeed we can suppose that we have the global integral constraint, $\CQ_{\Sigma} = 0$ in Newtonian cosmology, that has to be respected. As a result of this, also the other variables are known on the global scale: $a_{\Sigma} = a(t)$, i.e. $\mathfrak{a}_{\Sigma} =1$, given $\langle\delta\rangle_{\Sigma} = 0$, which is the definition of the background density, $\varrho_H := \langle\varrho\rangle_\Sigma$ (the averaged density contrast must vanish on the background globally matching initial data). Thus, in the splitting equations \eqref{volpartsyspec} the left-hand sides are known. 
Therefore, the knowledge of the global variables reduces the problem by two independent variables, so that we can solve the three-domain family of systems provided we impose \textit{one} further constraint.
For an $\ell$-fold partitioning there are $3\ell$ unknowns ($\mathfrak{a}^i_{\CD}$, $\CQ^i_{\CD}$, $\langle\delta\rangle^i_{\CD}$) against $2\ell +1$ independent relations 
($\ell$ continuity, $\ell$ acceleration, and $\sum_i{(\mathfrak{a}^i_{\CD})^3}=1$); hence the number of closure conditions needed is $\ell -1$.

One realization for $\ell =2$ would be to assume separate models for $M$- and/or $E$-regions.
As an example we take a dispersion-supported region in
Jeans equilibrium (\textit{cf.} \eqref{jeans} for $\bar{\bm v}=0$) as the $M$-region closure. We assume the stable clustering hypothesis, $H_M=0$, as the only needed functional closure condition. This  implies ${\ddot a}_M = 0$, i.e. the condition:
\begin{equation}
\label{Mclosure}
\CQ_M - \left(4\pi G\langle\varrho\rangle_M-\Lambda\right) = \frac{{\ddot a}_M}{a_M} = 0 \; .
\end{equation}
The global constraint $\CQ_\Sigma=0$ in turn fixes the void backreaction
through the last line of \eqref{volpartsys2}:
\begin{equation}
\CQ_E=-\frac{1-\lambda_E}{\lambda_E}\Big(4\pi G\langle\varrho\rangle_M-\Lambda\Big)
      -6\,(1-\lambda_E)\,H_E^{2}\; ,
\end{equation}
together with $\dot\lambda_E=3\lambda_E(1-\lambda_E)H_E$ from \eqref{volpart21} and
$\langle\varrho\rangle_M=\mathrm{const.}$ (since $a_M=\mathrm{const.}$). 
We infer that the $M$-region contribution enters
with the weight $(1-\lambda_E)/\lambda_E$, so it is progressively
switched off as voids come to dominate, but $\langle\varrho\rangle_M$ is expected to be large. For a virialized overdense region, for which $4\pi G\langle\varrho\rangle_M>\Lambda$, both contributions are negative and hence $\CQ_E<0$.

This example also makes transparent that plausible physical conditions on $M$-regions are subordinated to the Newtonian $3$-torus ``forcing condition'', i.e. that voids have to carry a decelerating backreaction to satisfy the global constraint.
We will comment more on this point in the discussion section~\ref{s-discussion}.

\section{Separate Universe conjecture}
\label{sec:separate}
The above-developed systematic volume partitioning with global constraints allows us to control the Newtonian problem on regional scales. In other words, this formalism allows us to make consistent assumptions on the behavior of regional spatial domains in the spirit of the \textit{Separate Universe Conjecture} without violating the global properties of Newtonian cosmologies.

The {\it Separate Universe Conjecture} (see e.g. \cite{Bernardeau}, \cite{separateuniverse}, \cite{carrharada}) states that under certain conditions, the initial over-density closely follows the spherical collapse model i.e. it behaves like a homogeneous model whose evolution is determined by the average density and positive spatial curvature. In its spherically-symmetric version this idea is closely related to the spherical collapse model, and usually requires a compensated density profile (top-hat). Within general relativity, the conformal version of Fermi coordinates \cite{conformalfermi} was used to state that locally, the isotropic part of a sufficiently long-wavelength perturbation can be absorbed into a local scale factor and curvature. The separate-universe idea has subsequently been used as a simplifying description of nonlinear structure formation and as a basis for regional cosmological models. A stronger assumption, however, is to approximate finite regions of an inhomogeneous Universe as independently evolving Friedmannian solutions, each equipped with its own local scale factor and mean density, while neglecting anisotropies, matching conditions, or constraints imposed by the global model. This latter assumption goes beyond the local separate-universe statement and requires additional consistency conditions when such regions are combined into a global cosmological model. 
Making restrictive assumptions on regional domains and considering the region to evolve separately from the ambient matter that is non-locally interacting with the considered domain, has to be put under careful scrutiny, especially in nonlinear modeling attempts. 

An example of such approach is the \textit{Average Expansion Rate Approximation (AvERA)}. In their paper \cite{nbody}, the authors calculate the change in the global expansion rate based on an extended form of \textit{Separate Universe Conjecture} described in the previous paragraph. Effectively, the authors use results of a Newtonian $N$-body simulation to calculate a volume increment in each of the simulation's sub-cubes using the Friedmann equations at each time step. Since the density contrasts depart from Gaussianity at the nonlinear stages of structure formation, the sum of volume increments from each cube is not equal to the global volume increment obtained from the global Friedmann equation:
\begin{equation}
V^{\rm FLRW}_{\Sigma}(t)\neq \sum_{\ell} V_{\ell}(t) \ .\nonumber
\end{equation}
There are two possible interpretations of the local scale factors and expansion rates entering the AvERA construction, and we will consider both. First, they may be regarded as approximations to the physical scale factors and expansion rates of regional domains, in the spirit of a separate Universe description. In this interpretation the regions should be identified with an actual spatial partition of the Newtonian model and are therefore subject to the exact partitioning relations. Alternatively, the local Friedmannian scale factors may be understood only as auxiliary statistical quantities assigned to the instantaneous density sampled in the simulation cells, without identifying them with the geometrical evolution of those cells or with Lagrangian matter domains. The latter interpretation does not permit a direct application of the Lagrangian partitioning equations, but the simulation cells can instead be treated as non-material control volumes, whose evolution is governed by the Reynolds transport theorem. We address these two interpretations separately below.\footnote{Another route is to impose statistical assumptions to construct effective equations; an early example is the assumption of homogeneous-isotropic turbulence by \cite{OlsonSachs}.}

If the local scale factors are treated within Lagrangian partitioning formalism, results obtained with AvERA contradict the $3$-torus constraint (which is a standard assumption in $N$-body simulations), see appendix~\ref{appB}. To see why \cite{nbody} possibly get a nonzero global backreaction from $3$-torus simulations, we recall \eqref{volpart8} for the global $3$-torus domain $\Sigma$,
\begin{eqnarray}
\CQ_{\Sigma} = \sum_{\ell} \lambda_{{\mathcal F}_{\ell}} \CQ_{{\mathcal F}_{\ell}} + 3\sum_{\ell\neq m} \lambda_{{\mathcal F}_{\ell}}\lambda_m(H_{{\mathcal F}_{\ell}} - H_m )^2 ~.
\label{globalbackreaction}
\end{eqnarray}
If, following \cite{nbody}, we put for all $\ell$, $\CQ_{{\mathcal F}_{\ell}}= 0$, we get an incorrect equation:
\begin{eqnarray}
\CQ_{\Sigma} =  3\sum_{\ell\neq m} \lambda_{{\mathcal F}_{\ell}}\lambda_m(H_{{\mathcal F}_{\ell}} - H_m )^2 ~,\nonumber
\end{eqnarray}
where the right-hand side is always positive and may mimic Dark Energy, in obvious contradiction with the global constraint
$\CQ_{\Sigma} =0$. This conclusion does not change if we include velocity dispersion (appendix~\ref{appB}) as an effective description of multi-streaming in shell-crossing, as speculated by the authors in \cite{compactnbody}, see also \cite{Kaiser17} and \cite{Buchert18}.
The statistical character of the AvERA construction does not, by itself, provide a justification for neglecting regional backreaction. Statistical homogeneity and isotropy constrain ensemble averages of the fields, but do not imply the vanishing of quadratic scalar invariants such as the variance of the expansion or the shear scalar. Whenever the regional expansion rates of an exact partitioning have a substantial variance, the volume-weighted regional backreaction is necessarily of the same magnitude and opposite sign. Increasing the number of statistically sampled regions does not remove this contribution.

One more subtlety regarding AvERA comes from the definition of sub-domains used in the code.
 The regular grid on which the local density is evaluated, is naturally interpreted as a partition into non-material control volumes. Matter may cross the boundaries of these cells and their density evolution is therefore governed by the Reynolds transport theorem \cite{Reynolds}, \cite{Flanders}, which is a generalization of the commutation rule in \eqref{eq:commutation}  (for the numerical investigation of the backreaction in non-material domains see e.g.  \cite{Koksbang}). The Reynolds theorem describes the behavior of a given scalar quantity, $\chi({\bf x},t)$, within an arbitrary control volume. Let us define a one-parameter set of local diffeomorphisms - $\{\phi_t\}$ and let $\bar{\CD}=\phi_t(\bar{\CD}_i)$ be the control volume, where $\bar{\CD}_i$ is the initial control volume, whose boundary moves with velocity ${\bf v}_{\rm b}$. We introduce the following definitions:
\begin{equation}
    \bar{V}_{\bar{\CD}}(t)=\int_{\bar{\CD}}{\rm d}^3x\;,\;a_{\bar{\CD}} = \left(\frac{\bar{V}_{\bar{\CD}}(t)}{\bar{V}_{\CD_i}}\right)^{1/3} \;,\;\langle \chi \rangle_{\bar{\CD}}:=\frac{1}{\bar{V}_{\bar{\CD}}}\int_{\bar{\CD}}\chi~{\rm d}^3x\;.
\end{equation}
It is also useful to introduce the relative velocity and the averaged flux of the scalar $\chi$ through the boundary:
\begin{equation}
    {\bf v}^{\rm rel} := \boldsymbol{\bar v} - {\bf v}_{\rm b}\;,\;{\mathcal F}_{\bar{\CD}}[\chi] := \frac{1}{\bar{V}_{\bar{\CD}}} \int_{\partial \bar{\CD}}\chi ~{\bf v}^{\rm rel}\cdot {\rm d}{\bf A}\;, 
\end{equation}
where  ${\rm d}{\bf A}$ is the surface element and we denote ${\mathcal F}_{\bar{\CD}}: = {\mathcal F}_{\bar{\CD}}[1]$.
Additionally, we have:
\begin{equation}
    3\bar{H}_{\bar{\mathcal D}} = \frac{\dot{\bar{V}}_{\bar{\mathcal D}}}{{\bar{V}}_{\bar{\mathcal D}}} = \langle \nabla \cdot  {\bf v}_{\rm b}\rangle_{\bar{\CD}}\;\;,\;\;\langle \theta \rangle_{\bar{\mathcal D}} =  3\bar{H}_{\bar{\mathcal D}}+{\mathcal F}_{\bar{\CD}}~.
    \label{cv-expansion}
\end{equation}

The Reynolds transport theorem reads:
\begin{equation}
    \frac{\rd}{\rd t} \int_{\bar{\CD}} \chi \rd^3x = \int_{\bar{\CD}} \left(\partial_t \chi + \nabla \cdot \left(\chi {\bf v}_{\rm b}\right)\right)\rd^3x \; ,
    \label{eq:reynolds}
\end{equation}
for an arbitrary boundary velocity $\bm v_{\rm b}$. If the boundary is material,
$(\bar{\bm v}-\bm v_{\rm b})\!\cdot\!\bm n=0$, one may replace
$\bm v_{\rm b}$ by $\bar{\bm v}$ and \eqref{eq:reynolds} reduces to
 \eqref{eq:commutation}. However, in general we write:
\begin{eqnarray}
\label{eq:reynolds_commutation}
\frac{\rd}{\rd t} \caverage{\chi} - \caverage{\ddt{\chi}} &=& \caverage{\chi \theta} - \caverage{\chi}\caverage{\theta} \nonumber \\ &-& \left({\mathcal F}_{\bar{\CD}}[\chi]-\caverage{\chi}{\mathcal F}_{\bar{\CD}}\right),
\end{eqnarray}
where the usual commutation formula is supplemented with non-zero flux terms.

The main consequence of the freedom of choice of the control volume is that in general the domains will not preserve the mass: 
\begin{eqnarray}
    &&\frac{\rm d}{\rm d t}\langle \varrho\rangle_{\bar{\CD}} +3\bar{H}_{\bar{\CD}}\caverage{\varrho} = -{\mathcal F}_{\bar{\CD}}[\varrho], \nonumber\\
    &&\frac{\rm d}{\rm d t}{M}_{\bar{\CD}} = -\bar{V}_{\bar{\CD}}{\mathcal F}_{\bar{\CD}}[\varrho] = -\int_{\partial \bar{\CD}} \varrho {\bf v}^{\rm rel}\cdot \rd {\bf A}\;.
\end{eqnarray} 
 We define a control-volume acceleration functional by writing the volume acceleration in Friedmann-like form:
 \begin{equation}
\label{avacceleration-R}
3\frac{\ddot{a}_{\bar{\CD}}}{a_{\bar{\CD}}} - \Lambda +4\pi G \caverage{\varrho} = \bar{\CQ}_{\bar{\CD}}\; ~,
 \end{equation}
where the backreaction term splits naturally into kinematical bulk and flux backreactions:
\begin{eqnarray}
    \bar{\CQ}_{\bar{\CD}} &=& \CQ_{\bar{\CD}}+\bar{\CQ}^{\rm flux}_{\bar{\CD}}, \nonumber \\
    \CQ_{\bar{\CD}} &=& \frac{2}{3}\left(\caverage{\theta^2}-\caverage{\theta}^2\right)+2\caverage{\omega^2-\sigma^2}-\caverage{\partial_i\left(\frac{\Psi_i}{\varrho}\right)}, \nonumber \\
    \bar{\CQ}^{\rm flux}_{\bar{\CD}} &=& -{\mathcal F}_{\bar{\CD}}[\theta]+\bar{H}_{\bar{\CD}}{\mathcal F}_{\bar{\CD}}+\frac{2}{3}{\mathcal F}^2_{\bar{\CD}} - \dot{{\mathcal F}}_{\bar{\CD}}~,
\end{eqnarray}
where $\CQ_{\bar{\CD}}$ is the backreaction term built from the fluid fields but averaged over the control volume. The algebraic partition identities for spatial averages remain valid for any disjoint partition at a given time, while the evolution equations of the regional averages acquire the Reynolds flux terms. In particular, if $\bar{\mathcal F}_l$ form a time-dependent partitioning, Eq.~\eqref{volpart8} becomes:
\begin{equation}\label{volpart_Rey}
    \bar{\CQ}_{\bar{\CD}} = \sum_{\ell} \bar{\lambda}_{\bar{{\mathcal F}}_{\ell}}\left( \CQ_{\bar{{\mathcal F}}_{\ell}}+\bar{\CQ}^{\rm flux}_{\bar{{\mathcal F}}_{\ell}} \right) + 3\sum_{\ell\neq \bar{m}} \bar{\lambda}_{\bar{{\mathcal F}}_{\ell}}\bar{\lambda}_{\bar{m}}(\bar{H}_{{\bar{\mathcal F}}_{\ell}} - \bar{H}_{\bar{m}} )^2 ~.
\end{equation}
On the other hand we have:
\begin{equation}
    \CQ_{\bar{\CD}} = \sum_{\ell} \bar{\lambda}_{{\bar{\mathcal F}}_{\ell}} \CQ_{\bar{{\mathcal F}}_{\ell}} + 3\sum_{\ell\neq {\bar{m}}} \bar{\lambda}_{\bar{{\mathcal F}}_{\ell}}\bar{\lambda}_{\bar{m}}(H_{{\bar{\mathcal F}}_{\ell}} - H_{\bar{m}} )^2 ~,
\end{equation}
where $H_{\bar{\mathcal F}_{\ell}} = \frac{1}{3}\langle \theta \rangle_{\bar{\mathcal F}_{\ell}}$. For the complete periodic domain, $\bar{\mathcal D}=\Sigma$, there is no external boundary flux and the global domain is material, hence $\mathcal F_\Sigma=0$, $\bar Q_\Sigma=Q_\Sigma$, and the Newtonian 3-torus constraint gives $Q_\Sigma=0$, leading to:
\begin{equation}
   \sum_{\ell} {\bar\lambda}_{\bar{\mathcal F}_{\ell}}\bar{\CQ}^{\rm flux}_{\bar{{\mathcal F}}_{\ell}}= 3\sum_{\ell\neq {\bar{m}}} \bar{\lambda}_{\bar{{\mathcal F}}_{\ell}}\bar{\lambda}_{\bar{m}}\left((H_{{\bar{\mathcal F}}_{\ell}} - H_{\bar{m}} )^2-(\bar{H}_{{\bar{\mathcal F}}_{\ell}} - \bar{H}_{\bar{m}} )^2\right)\; .
   \label{Q_flux}
\end{equation}
Although the net volume and mass fluxes across the internal boundaries cancel upon summation over the complete partition, the flux backreaction terms do not vanish separately. Rather, they compensate the difference between the variance of the matter-expansion rates and that of the control-volume expansion rates. Passing from material to non-material domains therefore does not remove the global constraint; it changes its regional decomposition.
Within the second interpretation of the AvERA construction,
the Reynolds formulation makes clear that the simulation cells themselves
cannot in general be regarded as independently evolving Friedmann domains.
The density is sampled in non-material control volumes, while the simulation
volume is rescaled with a common effective scale factor. 
Hence the geometrical expansion rate of all cells is
\begin{equation}
    \bar H_{\bar{\mathcal F}_{\ell}}=:H_{\rm eff}\; .
\end{equation}
This specialization makes the relation between kinematical and
flux backreaction particularly transparent. For a complete partition
of the periodic domain, the fluxes
${\mathcal F}_{\bar{\mathcal F}_{\ell}}
:={\mathcal F}_{\bar{\mathcal F}_{\ell}}[1]$
satisfy
\begin{equation}
   \sum_{\ell}
   \bar{\lambda}_{\bar{\mathcal F}_{\ell}}
  {\mathcal F}_{\bar{\mathcal F}_{\ell}}  = 0 \; ,
\end{equation}
since the contributions from shared internal interfaces cancel
pairwise and the complete $3$-torus has no external boundary.
Furthermore, Eq.~\eqref{cv-expansion} gives
\begin{equation}
    H_{\bar{\mathcal F}_{\ell}}
    =
    H_{\rm eff}
    +\frac{1}{3}
    {\mathcal F}_{\bar{\mathcal F}_{\ell}} \; .
\end{equation}
Consequently, Eq.~\eqref{Q_flux} reduces to
\begin{eqnarray}
    \sum_{\ell}
    \bar{\lambda}_{\bar{\mathcal F}_{\ell}}
    \bar{\CQ}^{\rm flux}_{\bar{\mathcal F}_{\ell}}
    &=&
    \frac{1}{3}
    \sum_{\ell \ne m}
    \bar{\lambda}_{\bar{\mathcal F}_{\ell}}
    \bar{\lambda}_{\bar{\mathcal F}_{m}}
    \left(
       {\mathcal F}_{\bar{\mathcal F}_{\ell}}
       -
       {\mathcal F}_{\bar{\mathcal F}_{m}}
    \right)^2
    \nonumber\\
    &=&
    \frac{2}{3}
    \sum_{\ell}
    \bar{\lambda}_{\bar{\mathcal F}_{\ell}}
    {\mathcal F}_{\bar{\mathcal F}_{\ell}}^{\,2}
    \geq 0 \; .
\end{eqnarray}
Since $\bar{\CQ}_{\Sigma}=\CQ_{\Sigma}=0$ and the variance of the
control-volume expansion rates vanishes, Eq. ~\eqref{volpart_Rey} equivalently yields:
\begin{equation}
    \sum_{\ell}
    \bar{\lambda}_{\bar{\mathcal F}_{\ell}}
    \CQ_{\bar{\mathcal F}_{\ell}}
    =
    -\frac{2}{3}
    \sum_{\ell}
    \bar{\lambda}_{\bar{\mathcal F}_{\ell}}
    {\mathcal F}_{\bar{\mathcal F}_{\ell}}^{\,2}
    \leq 0 .
\end{equation}
Thus, whenever the matter-expansion rates differ from the common
geometrical expansion rate of the cells, a positive volume-weighted
flux backreaction is necessarily accompanied by an equal and opposite
volume-weighted kinematical backreaction. 
Both contributions vanish
simultaneously only if
${\mathcal F}_{\bar{\mathcal F}_{\ell}}=0$ for every cell.

The common geometrical expansion rate does not, however, imply
individual mass conservation of the cells. Their density evolution is
instead governed by
\begin{equation}
    \frac{{\rm d}}{{\rm d}t}
    \left\langle \varrho \right\rangle_{\bar{\mathcal F}_{\ell}}
    + 3 H_{\rm eff}
    \left\langle \varrho \right\rangle_{\bar{\mathcal F}_{\ell}}
    =
    - {\mathcal F}_{\bar{\mathcal F}_{\ell}}[\varrho] \; .
\end{equation}
This differs from the dust continuity equation of an isolated Friedmann
domain,
\begin{equation}
    \dot{\varrho}_{\ell}
    + 3 H_{\ell}^{\rm SU} \varrho_{\ell} = 0 \; ,
\end{equation}
where we denote by
\begin{equation}
    H_{\ell}^{\rm SU}
    :=
    \frac{\dot a_{\ell}^{\rm SU}}{a_{\ell}^{\rm SU}}
\end{equation}
the auxiliary separate Universe expansion rate assigned to the $\ell$th
cell by solving a homogeneous-isotropic Friedmann model whose matter
density is identified, at the given time step, with the density measured
in that cell,
\begin{equation}
    \varrho_{\ell}^{\rm SU}
    :=
    \left\langle \varrho \right\rangle_{\bar{\mathcal F}_{\ell}} \; .
\end{equation}
If this identification is required to hold dynamically, rather than merely being imposed instantaneously at each time step, consistency would require
\begin{equation}
    {\mathcal F}_{\bar{\mathcal F}_{\ell}}[\varrho]
    =
    3
    \left\langle \varrho \right\rangle_{\bar{\mathcal F}_{\ell}}
    \left(
        H_{\ell}^{\rm SU} - H_{\rm eff}
    \right) ,
\end{equation}
a relation which is not implied by the density assignment alone.

The same issue appears in the acceleration law. The actual control-volume
dynamics obeys:
\begin{equation}
    3\frac{\ddot a_{\bar{\mathcal F}_{\ell}}}
            {a_{\bar{\mathcal F}_{\ell}}}
    - \Lambda
    + 4\pi G
      \left\langle \varrho \right\rangle_{\bar{\mathcal F}_{\ell}}
    =
    \bar Q_{\bar{\mathcal F}_{\ell}}
    =
    Q_{\bar{\mathcal F}_{\ell}}
    +
    \bar Q^{\rm flux}_{\bar{\mathcal F}_{\ell}} \; .
\end{equation}
Consequently, the regional evolution is not Friedmannian unless the combined
kinematical and flux backreaction vanishes. Since all simulation cells share
the same geometrical scale factor while their densities differ,
$\bar Q_{\bar{\mathcal F}_{\ell}}$ is generically non-zero.

The local Friedmann equations used in AvERA may therefore consistently be
regarded as defining an auxiliary statistical prescription,
\begin{equation}
    \left\langle \varrho \right\rangle_{\bar{\mathcal F}_{\ell}}
    \longmapsto
    H_{\ell}^{\rm SU} ,
\end{equation}
but not, without additional flux and backreaction conditions imposed, as required by 
the dynamical equations governing the actual simulation cells. The statistical
interpretation avoids identifying the auxiliary Friedmann regions with
geometrical subdomains, but it does not make the physical control-volume
backreaction-free or the mass-flux terms vanish.

In the framework of general relativity, such a global constraint on backreaction does not exist, but still one has to control constraints globally. We can now understand that one has to be cautious when concluding on non-vanishing global backreaction through simplifying assumptions on regional domains. Examples here include the wall-void model by \cite{W07}, \cite{W14}. 

Another, slightly different case is provided by the silent-virialization assumption \cite{boud18}. The imposed stable clustering phase may be interpreted within the present framework as replacing the continued collapse of a Lagrangian domain by a frozen, or nearly frozen, spatial region. This replacement is, however, not compatible in general with retaining the surrounding domains as a Lagrangian partition.

Indeed, the boundary of every neighbouring Lagrangian domain must move with the normal matter velocity,
\begin{equation}
\label{materialboundary}
(\boldsymbol{\bar v}-\mathbf v_b)\cdot\mathbf n=0 \; .
\end{equation}
The same interface is simultaneously the boundary of the virialized region. If this region is frozen geometrically, the boundary velocity is prescribed independently of the matter flow; for a strictly frozen physical boundary, $\mathbf v_b\cdot\mathbf n=0 
$.

Both conditions can hold on the same interface only in the special case
$\boldsymbol{\bar v}\cdot\mathbf n=0$.
Hence, in general, one cannot simultaneously keep the virialized region frozen, the neighbouring regions Lagrangian, and the collection of regions as an exact spatial partitioning. If the frozen boundary is retained, the adjacent regions cease to be material and their individual masses are no longer conserved. If the neighbouring regions are kept Lagrangian, the frozen region can no longer share their evolving boundaries and the original exact partitioning is lost.

At the integrated level, global mass conservation together with the mass conservation of all remaining Lagrangian domains would additionally require:
\begin{equation}
\dot M_{\bar{\mathcal F}_l}
=-\bar V_{\bar{\mathcal F}_l}
{\mathcal F}_{\bar{\mathcal F}_l}[\varrho]=0 \; .
\end{equation}
Thus, a zero net mass flux through the frozen region is a necessary condition for mass conservation, but it is not sufficient to restore the exact Lagrangian partitioning. Compatibility of the shared interfaces requires the stronger pointwise material boundary condition \eqref{materialboundary}.

\section{Discussion}
\label{s-discussion}

We investigated an exact and dynamical (Lagrangian) volume partitioning of inhomogeneous Newtonian cosmologies for self-gravitating collisionless matter. We also generalized known theorems when velocity dispersion is included. The Newtonian formulation allowed us to employ those theorems as global integral constraints on the multi-scale family of coupled dynamical equations. Similar theorems are limited to flat space sections in general relativity, for a summary see \cite{Buchert11Towards}. We so obtained a framework that allows looking at Newtonian simulations through assumptions on regional domains, while having control on their global properties.

We found that regional assumptions may contradict global constraints in Newtonian cosmology. As an example we showed that AvERA \cite{nbody} leads to spurious global backreaction. A word is in order here: first, to identify the main effect of backreaction with the difference in the Hubble rates between voids $E$ and walls $M$ \cite{W07}, \cite{W14} in Equation~(\ref{globalbackreaction}) is physically expected to be a dominating effect due to the volume-dominance of voids in the late Universe. However, second, the intuition that dispersion-supported $M$-regions as well as voids $E$ may both bear the physical tendency to become spherical and, hence, arguing for the vanishing of the regional backreaction terms to a good approximation, turns out to be misleading. Both (plausible) arguments apparently fail in Newtonian cosmology: advancing the argument of almost spherical voids, the wall-domains will be highly shear-dominated, while advancing the argument of `virialized', almost spherical $M$-regions, expected on very small scales, would not leave well-defined spherical voids.
However, and here we speculate, especially the argument of dominance of the difference in the Hubble rates between the two regions may hold the main effect and may turn out to be mainly misaligned due to the forcing constraint to a $3$-torus topology. It is this forcing constraint, being mandatory for the architecture of Newtonian cosmologies, that merits to be strongly questioned when we move to realistic cosmologies within general relativity. A similar forcing of variables to match a global $3$-torus architecture is observed elsewhere, e.g. in the reconstruction of variables from observed peculiar-velocity data \cite{Galoppo}: if we would set the cosmological constant to zero in the $3$-torus model, the reconstructed data would favour a dominant negative average curvature globally in the present-day Universe, which does not match an assumed flat background cosmology. 

Within general relativity this problem becomes interesting, since we expect global backreaction to occur, mainly because of emerging curvature \cite{buchert:dust}, \cite{Buchert08status}, \cite{BuchRas12}, \cite{Bolejko1,Bolejko2} together with the non-conservation of spatial curvature \cite{buchertcarfora}.
The multi-partitioning formalism is already implemented relativistically \cite{multiscale}. The Lagrangian models for the realization of structure formation have also been developed, where the Lagrangian perturbation gradient is generalized to Cartan one-form fields, see \cite{Universe} and references therein. $M$-regions have been investigated in \cite{DM}. A possible route to go is via Hodge decomposition of the one-form fields \cite{rza4} featuring an integrable part -- related to what we analyzed in this paper -- and a non-integrable part that gives rise to a global contribution. \\

\textbf{Acknowledgements:}
We wish to thank Martin Kerscher for useful discussions and valuable remarks on the manuscript.
This work is part of a project that has received funding from the European Research Council (ERC) under the European Union's Horizon 2020 research and innovation program (grant agreement ERC Advanced Grant 740021--ARTHUS (2017--2023), PI: TB). 
All equations were independently cross-checked with the assistance of
Anthropic's Claude (Opus 5); the authors take full responsibility for
the content.

\appendix

\renewcommand{\theequation}{\thesection.\arabic{equation}}
\setcounter{equation}{0}

\section{Velocity moments of Vlasov's equation}
\label{appA}

In this appendix we closely follow \cite{BinneyTremaine}.\footnote{This appendix is an excerpt of
lecture notes \cite{BuchertM2}. It is included to provide a self-contained presentation of the system we study. More extensive details can be found in \cite{Fardeau}.}

We start by considering the evolution equation for the phase space density $f$ in the Eulerian phase space with the independent variables ${\bf w}\equiv (\boldsymbol{x}, \boldsymbol{v})$ and $t$:
\begin{equation}\label{Evo1}
\frac{Df}{Dt} \equiv \frac{\partial f}{\partial t} + v_i\frac{\partial f}{\partial x_i} + b_i\frac{\partial f}{\partial v_i} = - f \boldsymbol{\nabla}_{\bf w}\cdot {\bf w}\quad ; \quad i,j,k = 1,2,3 \ .
\end{equation} 
This is a continuity equation reflecting the conservation of the restmass of a fluid element in phase space along the phase space flow $\bf w$. The acceleration $\boldsymbol{b}$ is not an independent variable and is assumed to be not explicitly velocity-dependent in Newton's theory, $b = \boldsymbol{b}(\boldsymbol{x}, t)$.
Therefore, the phase space flow becomes incompressible (the phase space divergence of $\bf w$ vanishes),
so that also the phase space volume is conserved. Specifying the forces, $ \boldsymbol{F} = m\boldsymbol{b} $, renders the kinematical equation (\ref{Evo1}) dynamical. We think of the acceleration as being solely caused by the gravitational field strength,
\begin{subequations}
\begin{equation}\label{bg}
    \boldsymbol{b} =: \boldsymbol{g}(\boldsymbol{x},t) ~.
\end{equation}
Equation (\ref{bg}) forms the content of Newton's equivalence principle that asserts the equality of inertial mass $m$ and gravitational mass $m_G$ in the Newtonian force balance condition:
\begin{equation}
  \boldsymbol{F} = m\boldsymbol{b} = m_G \boldsymbol{g} = \boldsymbol{F}_G  ~~;~~ m_G = m ~.    \end{equation}
\end{subequations}
The Newtonian gravitational force field is furthermore conservative, i.e. derived from a potential:
\begin{equation}\label{Poisson1}
    \boldsymbol{g} =: -\boldsymbol{\nabla} \Phi ~.
\end{equation}
We conclude that, for a self-gravitating (collisionless) continuous system,
the following evolution equation called the Vlasov equation holds:
\begin{equation}\label{Vlasov}
\frac{\partial{f}}{\partial{t}} +v_i \frac{\partial{f}}{\partial{x_i}} - \frac{\partial{\Phi}}{\partial{x_i}}\frac{\partial{f}}{\partial{v_i}} = 0
~. \end{equation}
To obtain evolution equations in space, we form velocity moments of $f$, providing a hierarchy of tensors:
\begin{subequations}
\begin{equation}
\varrho(x,t) \equiv m \int_{\Omega_t^v} f \mathrm{d}^3v~~;~~\varrho\bar{v}_i \equiv m \int_{\Omega_t^v}v_if\mathrm{d}^3 v ~~;~~\cdots \ .
\end{equation}
The restmass density is a scalar quantity, a tensor of rank $0$, the restmass of a fluid element is abbreviated by $m$. The restmass current density is a vector quantity (rank $1$); the second moment involves the average of the quadratic tensor $v_iv_j$ (rank $2$), 
\begin{equation}
\varrho \overline{v_iv_j} \equiv m \int_{\Omega_t^v} v_iv_jf\mathrm{d}^3 v \ ,
\end{equation}
\end{subequations}
and so on.
We now proceed by computing the moments of the Vlasov equation. First, we compute the zeroth moment:
\begin{equation}
m\int_{\Omega^t_v} \Big( \frac{\partial{f}}{\partial{t}} +v_i \frac{\partial{f}}{\partial{x_i}} + g_i\frac{\partial{f}}{\partial{v_i}} \Big) \,\mathrm{d}^3v = 0 \ ,
\end{equation}
with $\Omega^t_v$ the integration domain in phase space restricted to the velocity half-space. We obtain, by computing each term:
\begin{subequations}
\begin{align}
m\int_{\Omega^t_v} \frac{\partial{f}}{\partial{t}}  \mathrm{d}^3v &= m \frac{\partial}{\partial t} \int_{\Omega^t_v} f  \mathrm{d}^3v = \frac{\partial\varrho}{\partial t} ~; \\
m\int_{\Omega^t_v} v_i \frac{\partial{f}}{\partial{x_i}}  \mathrm{d}^3v &= m \frac{\partial}{\partial x_i} \int_{\Omega^t_v} v_i f   \mathrm{d}^3v = \frac{\partial}{\partial x_i} (\varrho \overline{v}_i) ~; \\
m\int_{\Omega^t_v} g_i\frac{\partial{f}}{\partial{v_i}}\mathrm{d}^3v &= mg_i \int_{\Omega^t_v} \frac{\partial{f}}{\partial{v_i}}\mathrm{d}^3v = mg_i \int _{\partial \Omega_t^v} f\cdot \mathrm{d}A_i^v = 0
~. \end{align}
\end{subequations}
The first two terms arise from the velocity moments obtained earlier, and the third term is obtained by using Gauss' theorem and requiring the flux of $f$ vanishes at the volume boundary. This requirement is ensured if the system is isolated, or if the velocity domain is large enough to bound the system.
We therefore obtain for the zeroth velocity moment:
\begin{equation}
\frac{\partial\varrho}{\partial t} + \frac{\partial}{\partial x_i} (\varrho \overline{v}_i) = 0
~, \end{equation}
provided we have control over the velocity half-space.

An analogous procedure is applied to the first velocity moment. We obtain:
\begin{equation}
m\int_{\Omega^t_v} v_i\Big( \frac{\partial{f}}{\partial{t}} +v_j \frac{\partial{f}}{\partial{x_j}} + g_j\frac{\partial{f}}{\partial{v_j}} \Big) \,\mathrm{d}^3v \; = \; 0
~. \end{equation}
We compute each of the three terms:
\begin{subequations}
\begin{align}
m\int_{\Omega^t_v} v_i\frac{\partial{f}}{\partial{t}}  \mathrm{d}^3v &= m \frac{\partial}{\partial t} \int_{\Omega^t_v} v_i f  \mathrm{d}^3v = \frac{\partial}{\partial t} (\varrho\overline{v}_i) ~; \\
m\int_{\Omega^t_v} v_iv_j \frac{\partial{f}}{\partial{x_j}}  \mathrm{d}^3v &= m \frac{\partial}{\partial x_j} \int_{\Omega^t_v} v_i v_j f   \mathrm{d}^3v = \frac{\partial}{\partial x_j} (\varrho \overline{v_iv_j}) ~; \\
m\int_{\Omega^t_v} v_i g_j\frac{\partial{f}}{\partial{v_j}}\mathrm{d}^3v &= mg_j \int_{\Omega^t_v} v_i\frac{\partial{f}}{\partial{v_j}}\mathrm{d}^3v \nonumber\\
&= mg_j \int_{\Omega^t_v} \Big (\frac{\partial{v_if}}{\partial{v_j}} - \delta_{ij}f \Big) \,\mathrm{d}^3v = -\varrho g_i
~. \end{align}
\end{subequations}
Here, the first two terms are also easy to compute thanks to previous moment calculations. In the third term we neglect the divergence term with the same argument as above. We finally obtain for the first moment of Vlasov's equation:
\begin{equation}
\frac{\partial}{\partial t} (\varrho\overline{v}_i) + \frac{\partial}{\partial x_j} (\varrho \overline{v_iv_j}) =\varrho g_i
~. \end{equation}
We introduce the reduced second moment tensor:
\begin{equation}
\Pi_{ij} \equiv \varrho \Big (\overline{v_iv_j}-\bar{v_i}\bar{v_j} \Big)
~. \end{equation}
We also introduce the Lagrangian derivative: ${\mathrm d}/{\mathrm d}t=\partial/\partial t +\overline{v}_i \partial/\partial x_i$. We finally obtain the Euler-Jeans equation:
\begin{equation}
\varrho\frac{\mathrm{d}}{\mathrm{d}t}\overline{v}_i = \varrho g_i - \frac{\partial}{\partial x_j}\Pi_{ij}
~. \end{equation}
In order to obtain an evolution equation for the reduced second moment tensor, we compute the second moment of the Vlasov equation:
\begin{equation}
m\int_{\Omega^t_v} v_iv_j\Big( \frac{\partial{f}}{\partial{t}} +v_k\frac{\partial{f}}{\partial{x_k}} + g_k\frac{\partial{f}}{\partial{v_k}} \Big) \,\mathrm{d}^3v \; = \; 0
~. \end{equation}
We again compute each of the three terms:
\begin{subequations}
\begin{align}
m\int_{\Omega^t_v} v_iv_j\frac{\partial{f}}{\partial{t}}  \mathrm{d}^3v &= m \frac{\partial}{\partial t} \int_{\Omega^t_v} v_iv_j f  \mathrm{d}^3v =  \frac{\partial}{\partial t} (\varrho\overline{v_iv_j}) ~; \\
m\int_{\Omega^t_v} v_iv_jv_k \frac{\partial{f}}{\partial{x_k}}  \mathrm{d}^3v &= m \frac{\partial}{\partial x_k} \int_{\Omega^t_v} v_i v_jv_k f   \mathrm{d}^3v = \frac{\partial}{\partial x_k} (\varrho \overline{v_iv_jv_k}) ~; \\
m\int_{\Omega^t_v} v_iv_j g_k\frac{\partial{f}}{\partial{v_k}}\mathrm{d}^3v &= mg_k \int_{\Omega^t_v} v_iv_j\frac{\partial{f}}{\partial{v_k}}\mathrm{d}^3v \nonumber\\&= mg_k \int_{\Omega^t_v} \Big (\frac{\partial{v_iv_jf}}{\partial{v_k}} - \delta_{ik}v_jf -\delta_{jk}v_if \Big) \,\mathrm{d}^3v \nonumber \\
&= -\varrho (g_i\bar{v_j}+ g_j\bar{v_i})
~. \end{align}
\end{subequations}
We used again the argument that the velocity distribution is limited on $\Omega_t^v$. We then get an equation that links the different velocity moments:
\begin{equation}
\frac{\partial}{\partial t}(\varrho\overline{v_iv_j}) + \frac{\partial}{\partial x_k} (\varrho \overline{v_iv_jv_k}) = \varrho (g_i\bar{v}_j + g_j\bar{v}_i)
~. \end{equation}     
We write the Lagrangian derivative of $\Pi_{ij}$:
\begin{align}
\frac{\mathrm{d}}{\mathrm{d}t} \Pi_{ij} &= \frac{\partial}{\partial t}\Pi_{ij} +\bar{v}_k\frac{\partial}{\partial x_k}\Pi_{ij} \nonumber ~;\\
&= - \frac{\partial}{\partial x_k} (\varrho \overline{v_iv_jv_k}) +  \varrho (g_i\bar{v}_j + g_j\bar{v}_i)- \frac{\partial}{\partial t}(\varrho\bar{v}_i\bar{v}_j) +\bar{v}_k\frac{\partial}{\partial x_k}\Pi_{ij} \nonumber\\
&= -\frac{\partial}{\partial x_k} L_{ijk}- \frac{\partial}{\partial x_k}(\varrho\bar{v}_i\bar{v}_j\bar{v}_k) - \frac{\partial}{\partial x_k} [\bar{v}_i\Pi_{jk} + \bar{v}_j\Pi_{ki} + \bar{v}_k\Pi_{ij}]\nonumber\\&\quad+  \varrho (g_i\bar{v}_j + g_j\bar{v}_i)- \frac{\partial}{\partial t}(\varrho\bar{v}_i\bar{v}_j) +\bar{v}_k\frac{\partial}{\partial x_k}\Pi_{ij}
~. \end{align}
We then define the reduced third moment tensor:
\begin{align}\label{defL}
L_{ijk} &\equiv \varrho\overline{(v_i-\bar{v}_i)(v_j-\bar{v}_j)(v_k-\bar{v}_k)} \nonumber\\&= \varrho(\overline{v_iv_jv_k} - \bar{v}_i\bar{v}_j\bar{v}_k) - [\bar{v}_i\Pi_{jk}+\bar{v}_j\Pi_{ki} + \bar{v}_k\Pi_{ij} ]
~. 
\end{align}
Then we simplify this result using the equations for the moments obtained earlier to finally obtain:
\begin{equation}
\frac{\mathrm{d}}{\mathrm{d}t} \Pi_{ij} = -\frac{\partial}{\partial x_k}L_{ijk} - \Big[\frac{\partial \bar{v}_i}{\partial x_k}\Pi_{jk} + \frac{\partial \bar{v}_j}{\partial x_k}\Pi_{ki} + \frac{\partial\bar{v}_k}{\partial x_k}\Pi_{ij}\Big]
~. \end{equation}
We thus obtain the first members of a hierarchy of equations which is composed of the continuity equation, the Euler-Jeans equation,  the evolution equation for the reduced second moment, and the gravitational field equations:
\begin{equation}\label{ENJ1app}
    \begin{aligned}
 \frac{\rd}{\rd t}\varrho &= - \varrho \bar{v}_{j,j}~;\\
 \frac{\rd}{\rd t}\bar{v}_i & = g_i  - \frac{1}{\varrho} \Pi_{ij,j}~;\\
\frac{\mathrm{d}}{\mathrm{d}t} \Pi_{ij} &= -L_{ijk,k} - [\bar{v}_{i,k}\Pi_{jk} + \bar{v}_{j,k}\Pi_{ki} + \bar{v}_{k,k}\Pi_{ij}]~;\\ 
 g_{j,i} - g_{i,j} &= 0 ~;\\
g_{k,k} &= \Lambda - 4 \pi G \varrho~.
    \end{aligned}
    \end{equation}
We call the above system the \textit{Euler-Newton-Jeans system} when dropping the third reduced velocity moment.

\section{Known theorems and their generalization to include velocity dispersion}
\label{appB}

\setcounter{equation}{0}
The following theorems without explicit reference to velocity dispersion are summarized in Sect.7 of \cite{Buchert11Towards}. 

\noindent The kinematical backreaction $\CQ_\CD$ appears in the general acceleration law \eqref{avacceleration} and is defined in \eqref{Q}:
\begin{equation}\label{QD1}
    \begin{aligned}
 \CQ_\CD &= 3\frac{{\ddot a}_\CD(t)}{a_\CD(t)} - \Lambda +4\pi G \langle\varrho\rangle_\CD ~;\\
 \CQ_\CD &= 2\langle\inII(\boldsymbol{v})\rangle_\CD -\frac{2}{3}\langle\inI(\boldsymbol{v})\rangle_\CD^2 -\langle \frac{\partial }{\partial x_i}\Big( \frac{1}{\varrho}\Psi_i \Big)\rangle_\CD~.\\
    \end{aligned}
    \end{equation}
The requirement $\CQ_\CD = 0$ on a given domain $\CD$ is necessary and sufficient for $a_\CD (t)$ to obey a Friedmannian acceleration law with domain-dependent mass $M_\CD$ and curvature constant $k_\CD$. Using \eqref{invariants}, we can write the backreaction term in terms of integrals of divergences:
\begin{align}
    \CQ_\CD &= \frac{1}{V_\CD}\int_\CD 2\inII(\boldsymbol{v}) \;\rd ^3x - \frac{2}{3}\Big(\frac{1}{V_\CD}\int_\CD \inI(\boldsymbol{v}) \;\rd^3x\Big)^2 \nonumber\\ &-\frac{1}{V_\CD}\int_\CD \frac{\partial }{\partial x_i}\Big( \frac{1}{\varrho}\Psi_i \Big)\;\rd^3x \nonumber\\
    &= \frac{1}{V_\CD}\int_\CD \boldsymbol{\nabla} \cdot\Big( \boldsymbol{v} \boldsymbol{\nabla} \cdot \boldsymbol{v}-\boldsymbol{v} \cdot\boldsymbol{\nabla} \boldsymbol{v}\Big)\;\rd^3x - \frac{2}{3}\Big(\frac{1}{V_\CD}\int_\CD \boldsymbol{\nabla}\cdot\boldsymbol{v}\,\rd^3x\Big)^2 
    \nonumber\\
    &-\frac{1}{V_\CD}\int_\CD \boldsymbol{\nabla} \cdot \Big( \frac{1}{\varrho}\boldsymbol{\Psi} \Big)\; \rd^3x \nonumber\\
    &= \frac{1}{V_\CD}\int_{\partial \CD}\Big( \boldsymbol{v} \boldsymbol{\nabla} \cdot \boldsymbol{v}-\boldsymbol{v} \cdot\boldsymbol{\nabla} \boldsymbol{v}\Big)\cdot \boldsymbol{\rd A} - \frac{2}{3V_\CD^2}\Big(\int_{\partial\CD}\boldsymbol{v}\cdot \boldsymbol{\rd A}\Big)^2 \nonumber\\&-\frac{1}{V_\CD}\int_{\partial\CD}\frac{1}{\varrho}\boldsymbol{\Psi}\cdot \boldsymbol{\rd A}~,
\end{align}
with $\boldsymbol{\rd A} = {\bf n}\,\rd A$, where $\bf n$ is the outward pointing normal to the boundary and $\rd A$ the boundary surface element. The last equation was obtained using Gauss' theorem. We can conclude that, for an isolated system, the backreaction $\CQ_\CD$ would vanish and the domain would expand like an irrotational homogeneous-isotropic distribution of matter (for the anisotropic and rotational cases see the theorem at the end of this appendix).

\bigskip

\noindent We can go one step further and prove the vanishing of the backreaction term for a non-isolated but spherically symmetric systems, known as \textit{Newton's Iron Spheres theorem}. We suppose a spherically symmetric mean velocity field on a spherical domain $\mathcal{B}$ with radius $r = r({\mathcal R}, t)$. First, we consider the special case without velocity dispersion, $\langle \frac{\partial }{\partial x_i}\Big( \frac{1}{\varrho}\Psi_i \Big)\rangle_\CD = 0$, and we denote the dust part of backreaction $\CQ^{\rm dust}_{\mathcal{B}}$ during this calculation. Using the fact that the average velocity field inside $\mathcal{B}$ is only depending on the radial distance to the origin and is always parallel to the radial unit vector $\boldsymbol{e}_r$, we have:
\begin{equation}\label{NIST1}
    \boldsymbol{v} = v(r)\boldsymbol{e}_r \ .
\end{equation}
We instantaneously have that the curl of $\boldsymbol{v}$ is vanishing:
\begin{align}\label{rotV}
    \boldsymbol{\nabla}\times \boldsymbol{v} &= \frac{1}{r\sin\theta}\Big(\frac{\partial}{\partial \theta}(\sin \theta v_\varphi) - \frac{\partial v_\theta}{\partial \varphi}\Big)\;  \boldsymbol{e}_r \nonumber\\+ &\Big(\frac{1}{r\sin\theta}\frac{\partial v_r}{\partial \varphi} - \frac{1}{r}\frac{\partial}{\partial r}(rv_\varphi)\Big)\;\boldsymbol{e}_\theta + \frac{1}{r}\Big(\frac{\partial}{\partial r}(rv_\theta) - \frac{\partial v_r}{\partial \theta}\Big) \boldsymbol{e}_\varphi = 0 \, .
\end{align}
Since spherical symmetry implies ${\bf v}=v(r){\bf e}_r$ and
${\bf v}=\nabla S$, we have
\begin{equation}
    {\bf n}={\bf e}_r, \qquad\nabla S=\frac{\partial S}{\partial r}\,{\bf n}=v(r)\,{\bf n},
\label{eq:spherical_normal}
\end{equation}
where ${\bf n}$ denotes the outward-pointing unit normal to
$\partial{\mathcal B}$.
We use this to compute the average of the first principal scalar invariant $\langle\inI(\boldsymbol{v})\rangle_{\mathcal{B}}$ through the velocity potential using Gauss' theorem:
\begin{equation}\label{NIST2}
 \langle\inI(\boldsymbol{v})\rangle_{\mathcal{B}} =\frac{3}{4\pi r^3} \int_{\mathcal{B}} \rd^3x \boldsymbol{\nabla}\cdot\boldsymbol{v} = \frac{3}{4\pi r^3}\int_{\partial\mathcal{B}}v(r)\boldsymbol{e}_r\cdot\boldsymbol{\rd A} = 3 \frac{v(r({\mathcal R},t))}{r({\mathcal R},t)}.
\end{equation}
Then, we use the same mathematical ansatz to compute $\langle\inII(\boldsymbol{v})\rangle_{\mathcal{B}}$ using also Gauss' theorem:
\begin{align}\label{NIST3}
    \langle\inII(\boldsymbol{v})\rangle_{\mathcal{B}} &= \frac{3}{4\pi r^3}\int_{\mathcal{B}} \rd^3x \frac{1}{2} \boldsymbol{\nabla} \cdot\Big( \boldsymbol{v} \boldsymbol{\nabla} \cdot \boldsymbol{v}-\boldsymbol{v} \cdot\boldsymbol{\nabla} \boldsymbol{v}\Big)\nonumber \\
    &= \frac{3}{4\pi r^3}\int_{\partial\mathcal{B}}\frac{v(r)}{2}\Big(\boldsymbol{e}_r(\boldsymbol{\nabla}\cdot v(r)\boldsymbol{e}_r)-(\boldsymbol{e}_r\cdot\boldsymbol{\nabla})v(r)\boldsymbol{e}_r\Big)\cdot \boldsymbol{\rd A}\nonumber \\
    &= \frac{3v^2(r({\mathcal R},t))}{r^2({\mathcal R},t)} = \frac{1}{3}\langle\inI(\boldsymbol{v})\rangle_{\mathcal B}^2 ~.
\end{align}
We see that the dust part of the backreaction functional \eqref{QD1}, $\CQ^{\rm dust}_{\mathcal B}$, is vanishing under those hypotheses. Therefore $a_{\mathcal B}(t)$ is following Friedmann's equation for dust. 

We now work out the remaining part of the backreaction functional, $\langle \frac{\partial }{\partial x_i}\Big( \frac{1}{\varrho}\Psi_i \Big)\rangle_\CD \neq 0$. 
This remaining part, denoted by $\CQ^{\boldsymbol{\Psi}}_{\mathcal B}$ is already written as a flux term through the boundary $\partial {\mathcal B}$:
\begin{equation}\label{NIST21}
  \CQ^{\boldsymbol{\Psi}}_{\mathcal B} = -\frac{3}{4\pi r^3}\int_{\partial {\mathcal B}} \frac{1}{\varrho} \boldsymbol{\Psi} \cdot \boldsymbol{\rd A}  ~.
\end{equation}
Under the spherical symmetry hypothesis, we can write $\boldsymbol{\Psi}$ as
\begin{equation}\label{NIST2hyp}
    \boldsymbol{\Psi} = \Psi(r)\boldsymbol{e}_r ~;
\end{equation}
we have:
\begin{equation}\label{NIST22}
    \CQ^{\boldsymbol{\Psi}}_{\mathcal B} = \frac{-3}{4\pi r^3}\int_{\partial {\mathcal B}} \frac{1}{\varrho(r)}\Psi(r)\boldsymbol{e}_r \cdot \boldsymbol{{\rm d}A}  = \frac{-3}{\varrho(r({\mathcal R},t))}\frac{\Psi (r({\mathcal R},t))}{r({\mathcal R},t)} ~.
\end{equation}
The above integral can be evaluated at the boundary of the sphere, say $r = \mathcal R$. 
In stellar systems, the assumption is that pressure terms and their derivatives (here the dispersion term) vanish at the surface of the star. This situation we may assign to an Iron Sphere conjecture for isolated systems with dispersion. 
In the most general case, if we ask for situations with vanishing backreaction, we face the following condition:
\begin{equation}\label{NIST23}
    |\,\Psi({\mathcal R})\,| \ll {\mathcal R}\,\varrho{(\mathcal R)} \; H_{\mathcal B}^2~.
\end{equation}
We could interpret this in terms of an asymptotic fall-off condition of a given density profile yielding an approximate vanishing of the backreaction for specific density profiles.

Another possibility to evaluate the velocity-dispersion contribution is to use
the decomposition introduced in \eqref{anisPI},
\begin{equation}
  \Pi_{ij} = p\,\delta_{ij} + \Delta_{ij} \;,
\end{equation}
with $\Delta_{ij}$ the anisotropic part
of \eqref{anisPI}; the relations below do not assume $|\Delta_{ij}|\ll p$. Under
spherical symmetry the divergence of the velocity-dispersion tensor is purely
radial,
\begin{equation}
  \Psi_r = \frac{\partial p}{\partial r}
  + \frac{\partial\Delta_{rr}}{\partial r}
  + \frac{1}{r}\Big(2\Delta_{rr}-\Delta_{\theta\theta}-\Delta_{\varphi\varphi}\Big)
  \;=\; \frac{\partial p_r}{\partial r} + \frac{2}{r}\big(p_r - p_t\big) \;,
  \label{eq:Psi_spherical}
\end{equation}
where the last form uses $\Delta_{\theta\theta}=\Delta_{\varphi\varphi}$,
which holds for strict spherical symmetry, and introduces the radial and
tangential pressures $p_r := p+\Delta_{rr}$, $p_t := p+\Delta_{\theta\theta}$.
By \eqref{NIST22}, the dispersion contribution to the backreaction of a spherical
domain $\mathcal{B}$ of radius $\mathcal{R}$ is then given entirely by the
boundary value of the dispersion force per unit mass,
\begin{equation}
  \CQ^{\Psi}_{\mathcal{B}}
  = -\,\frac{3}{\mathcal{R}}\,
  \frac{\Psi_r(\mathcal{R})}{\varrho(\mathcal{R})} \;.
  \label{eq:Qpsi_spherical}
\end{equation}
Using the equation of state obtained in Equation~\eqref{eqstate1},
$p(r)=\kappa(r)\varrho^{5/3}(r)$, this reads explicitly:
\begin{eqnarray}
  \CQ^{\Psi}_{\mathcal{B}}
  &=& -\frac{3}{\mathcal{R}}\left[\,
  \frac{5}{3}\,\frac{\kappa(\mathcal{R})}{\varrho^{1/3}(\mathcal{R})}
  \frac{\partial\varrho}{\partial \mathcal{R}}
  + \varrho^{2/3}(\mathcal{R})\frac{\partial\kappa}{\partial \mathcal{R}}
  \right. \nonumber\\
  && \left. \quad
  + \frac{1}{\varrho(\mathcal{R})}\frac{\partial\Delta_{rr}}{\partial \mathcal{R}}
  + \frac{2}{\mathcal{R}\,\varrho(\mathcal{R})}
  \Big(\Delta_{rr}-\Delta_{\theta\theta}\Big)_{\mathcal{R}}\,\right] \;.
  \label{eq:Qpsi_spherical_eos}
\end{eqnarray}
The dispersion contribution therefore vanishes if and only if the radial
dispersion force per unit mass vanishes at the boundary,
\begin{equation}
  \left.\frac{\Psi_r}{\varrho}\right|_{\mathcal{R}} = 0 \;,
  \label{eq:Qpsi_zero}
\end{equation}
provided the corresponding boundary quantities are regular. In the isotropic
limit, $\Delta_{ij}=0$, \eqref{eq:Qpsi_spherical_eos} reduces to
\begin{equation}
  \CQ^{\Psi}_{\mathcal{B}}
  = -\,\frac{5\,\kappa(\mathcal{R})}{\mathcal{R}\,\varrho^{1/3}(\mathcal{R})}
  \frac{\partial\varrho}{\partial \mathcal{R}}
  \;-\; \frac{3\,\varrho^{2/3}(\mathcal{R})}{\mathcal{R}}
  \frac{\partial\kappa}{\partial \mathcal{R}} \;,
  \label{eq:Qpsi_isotropic}
\end{equation}
so that the condition \eqref{eq:Qpsi_zero} becomes
$\partial p/\partial\mathcal{R}=0$, and, for $\kappa=\mathrm{const.}$, simply
$\partial\varrho/\partial\mathcal{R}=0$ at the boundary radius.

Hence the vanishing of the pressure itself at the boundary is not, in general,
sufficient for $\CQ^{\Psi}_{\mathcal{B}}=0$; the relevant condition concerns
the radial dispersion force per unit mass at the boundary. The hydrostatic case makes this explicit. For the non-streaming,
spherically symmetric Jeans equilibrium considered here,
$\boldsymbol{\bar v}=0$, and hence
$\theta=0$, $H_{\mathcal B}=0$, and
$\dot a_{\mathcal B}=0$. Stationarity is therefore already imposed
at the kinematical level. Using
$\Psi_r/\varrho=\boldsymbol{g}\cdot\boldsymbol{e}_r$,
Gauss' theorem gives:
\begin{equation}
    \frac{g_r({\mathcal R})}{\mathcal R}
    =
    \frac{\Lambda}{3}
    -
    \frac{4\pi G}{3}
    \langle\varrho\rangle_{\mathcal B}\; .
\end{equation}
Equation~\eqref{eq:Qpsi_spherical} then yields:
\begin{equation}
\label{jeans}
    \CQ_{\mathcal B}
    =
    \CQ_{\mathcal B}^{\Psi}
    =
    4\pi G\langle\varrho\rangle_{\mathcal B}
    -
    \Lambda \; .
\end{equation}
The content of this relation is therefore not only the vanishing of the
regional acceleration $\ddot{a}_{\mathcal B}$, which already follows from stationarity, but
the value of the regional backreaction required to balance
self-gravity. A dispersion-supported stationary region thus carries,
in general, a non-vanishing regional backreaction.

A further global integral constraint arises in Newtonian models and N-body simulations that are subjected to periodic boundary conditions. For this purpose we have to use the construction in Hubble-comoving coordinates presented in section~\ref{Hubblecomoving}. For periodic fields off a Friedmannian background cosmology we think of a global domain $\Sigma$ as being a box whose opposite faces are identified. The domain would be topologically equivalent to a $3-$torus having no boundary and, hence, flux terms across the boundary would vanish. 
We introduce a peculiar-velocity field $\boldsymbol{u}$ and split the principal scalar invariants of its gradient as we did in \eqref{invcom2}. 
Writing naively $\boldsymbol{\Psi} = \boldsymbol{\Psi}_H + \boldsymbol{\Psi}^{\rm pec}({\bf q},t)$, we realize that $(\Psi_H)_i = \partial_j (p_H\,\delta_{ij}) = 0$, so that 
$\boldsymbol{\Psi}^{\rm pec} = \boldsymbol{\Psi}$, and the full backreaction term becomes:
\begin{equation}\label{invcom3}
    \CQ_\CD = 2\langle\inII(\tfrac{\boldsymbol{u}}{a})\rangle_{\CD} -\frac{2}{3}\langle\inI(\tfrac{\boldsymbol{u}}{a})\rangle^2_{\CD} - \frac{1}{a}\Big\langle \frac{\partial }{\partial q_i}\Big( \frac{1}{\varrho}\Psi_{i} \Big)\Big\rangle_{\CD}~.
\end{equation}
It is then solely determined through the inhomogeneities encoded in the deviations $\boldsymbol{u}$ from the Hubble field $\boldsymbol{v}_H$. Using these results, we may assume that a global Hubble flow exists on a large scale on which periodic boundary conditions can be imposed on $\boldsymbol{u}$. The universe model is then composed of compilations of a periodic box, with a Friedmannian scale factor describing the global evolution, $a_\Sigma(t) = a(t)$, since the global value of backreaction vanishes. This is proven by noting that 
the backreaction term \eqref{invcom3} can be written in terms of boundary terms,
\begin{align}\label{incom32}
\CQ_\CD &= \frac{1}{V_\CD}\int_{\partial \CD}\Big( \boldsymbol{u} \boldsymbol{\nabla} \cdot \boldsymbol{u}-\boldsymbol{u} \cdot\boldsymbol{\nabla} \boldsymbol{u}\Big)\cdot \boldsymbol{\rd A} - \frac{2}{3V_\CD^2}\Big(\int_{\partial\CD}\boldsymbol{u}\cdot \boldsymbol{\rd A}\Big)^2 \nonumber\\
&-\frac{1}{V_\CD}\int_{\partial\CD}\frac{1}{\varrho}\boldsymbol{\Psi}\cdot\boldsymbol{\rd A}~.
\end{align}
For a $3-$torus domain $\Sigma$ the boundary $\partial\Sigma$ is empty and we have proven a generalization of the Buchert-Ehlers theorem including velocity dispersion \cite{BuchertEhlers97,Buchert18}.
However, on smaller scales, fluctuations in the fields within the global domain would influence the dynamics
inside the universe model and may become important on smaller scales (see \cite{BKS}).

The following is a more general expansion law (for dust given in \cite[Eq. (B4)]{BuchertEhlers97} that we here write with inclusion of the multistream force term:
\begin{align}
\label{general}
3\frac{{\ddot a}_{\CD}}{a_{\CD}} + 4\pi G \frac{M_{\CD}}{V_i a_{\CD}^3 } -
\Lambda =  {\CQ}_{\CD} = \qquad\qquad\qquad\nonumber\\ 
\langle\boldsymbol{\nabla}\cdot \lbrack {\mathbf u} (\boldsymbol{\nabla} \cdot {\mathbf u} ) -
({\mathbf u} \cdot \boldsymbol{\nabla}) {\mathbf u} \rbrack \rangle_{\CD}
- \frac{2}{3} \langle\boldsymbol{\nabla} \cdot {\mathbf u} \rangle_{\CD}^2 - \Big\langle \boldsymbol{\nabla}\cdot \Big( \frac{1}{\varrho}\boldsymbol{\Psi}\Big)\Big\rangle_\CD
\nonumber\\
+2\left((\Omega^{\rm gb})^2 - (\Sigma^{\rm gb})^2 \right) +
2\left(\Omega_{ij}^{\rm gb}\langle\hat{\omega}_{ij}\rangle_{\CD}
- \Sigma^{\rm gb}_{ij}\langle\hat{\sigma}_{ij}\rangle_{\CD}\right).\qquad\qquad
\end{align}
Above, the peculiar-velocity's shear and vorticity are denoted by $\boldsymbol{\hat{\sigma}}$ and $\boldsymbol{\hat{\omega}}$,
respectively. Omitting global shear $\boldsymbol{\Sigma}^{\rm gb}$ and global vorticity $\boldsymbol{\Omega}^{\rm gb}$ of the model universe, we 
come back to the general expansion law on a homogeneous-isotropic Friedmannian background.\\

\noindent{\bf Remark:} According to the logical line of arguments for proving the above theorems, we can easily generalize these theorems by including further physical sources like pressure, deviations from mean field gravity and others \cite{Buchert05adhesive}. These sources would all appear in the backreaction functional. It remains then to be proven that these extra source terms are divergences. If they are not, we would obtain a global backreaction for $3$-torus cosmologies. We note that Newton's theory is a vector theory and the backreaction is a scalar functional. Said this, we can expect that source terms depend on principal scalar invariants of the peculiar-velocity gradient and, due to the vector character of the theory, can be integrated to flux terms.
This remark not only applies to Newtonian models, but also to general-relativistic models that set up deviations  off a Friedmannian cosmology through integrable fields on flat space sections with periodic boundary conditions.

\end{document}